\documentclass[superscriptaddress,preprintnumbers,amssymb,nofootinbib]{revtex4-2}

\pdfoutput=1 

\usepackage[T1]{fontenc}
\usepackage{graphicx}
\usepackage{epsfig}
\usepackage{amsmath}
\usepackage{amsfonts}
\usepackage{amssymb}
\usepackage{braket}
\usepackage{cancel}
\usepackage{pstricks}
\usepackage{color}
\usepackage{feynmf}
\usepackage[normalem]{ulem}
\usepackage{epstopdf}

\def\lsim{\:\raisebox{-0.5ex}{$\stackrel{\textstyle<}{\sim}$}\:}
\def\gsim{\:\raisebox{-0.5ex}{$\stackrel{\textstyle>}{\sim}$}\:}

\def\missET {{\not\!\! E_T}}

\usepackage{subcaption}
\usepackage{braket}
\usepackage{cancel}
\usepackage{setspace}
\usepackage{pstricks}
\usepackage{xcolor}
\usepackage{float}
\usepackage{mathtools}
\usepackage{booktabs}
\usepackage{enumitem}

\providecommand{\tabularnewline}{\\}

\def\gsim{\lower0.5ex\hbox{$\:\buildrel >\over\sim\:$}}
\def\lsim{\lower0.5ex\hbox{$\:\buildrel <\over\sim\:$}}

\def \lp{{L\!\!\!/}}

\begin{document}

\title{R-Parity Violating Supersymmetry and the 125 GeV Higgs signals}

\author{Jonathan Cohen,}
\email{jcohen@campus.technion.ac.il}
\author{Shaouly Bar-Shalom,}
\email{shaouly@physics.technion.ac.il}
\author{Gad Eilam,}
\email{eilam@physics.technion.ac.il}
\affiliation{Physics Department, Technion--Institute of Technology, Haifa 3200003, Israel}
\author{Amarjit Soni}
\email{adlersoni@gmail.com}
\affiliation{Physics Department, Brookhaven National Laboratory, Upton, NY 11973, US}

\date{\today}



\begin{abstract}
We study the impact of R-parity violating Supersymmetry (RPV SUSY) 
on the 125 GeV Higgs production and decay modes at the LHC. 
We assume a heavy SUSY spectrum with multi-TeV squarks and SU(2) scalar singlets as well as 
the decoupling limit in the SUSY Higgs sector. In this case 
the lightest CP-even Higgs is SM-like when R-parity is conserved. In contrast, we show that when R-parity violating interactions are added to the SUSY framework, significant deviations may occur in some production and decay channels of the 125 GeV Higgs-like state. 
Indeed, we assume a single-flavor 
(mostly third generation) 
Bilinear RPV (BRPV) interactions, which generate Higgs-sneutrino mixing, lepton-chargino mixing and 
neutrino-neutralino mixing, and find that notable 
deviations of ${\cal O}(20-30\%)$  may be expected in the Higgs signal strength observables in some channels, e.g., in $p p \to h \to \mu^+ \mu^-, \tau^+ \tau^-$. 
Moreover, we find that new and detectable signals 
associated with BRPV Higgs decays to gauginos, $h\rightarrow\nu_{\tau}\tilde{\chi}_{2}^{0}$ and 
    $h\rightarrow\tau^{\pm}\chi_{2}^{\mp}$, may also arise in this scenario. These decays yield a typical signature of $h \to \tau^\pm \ell^\mp + \missET$ ($\ell =e,\mu,\tau$) that can be much larger than in the SM, and may also be accompanied by an ${\cal O}(20-30\%)$ enhancement in the di-photon signal $pp \to h \to \gamma \gamma$. 
We also examine potential interesting signals of Trilinear 
R-parity violation (TRPV) interactions in the production and decays of the Higgs-sneutrino BRPV mixed 
state (assuming it is the 125 GeV scalar) and show that, in this case also, large deviations up to ${\cal O}(100\%)$ 
are expected in e.g., $p p \to h \to \mu^+ \mu^-, \tau^+ \tau^-$, which are sensitive to the BRPV$\times$TRPV coupling product.
\end{abstract}


\maketitle
\flushbottom

\section{\label{sec:Introduction}Introduction}

\pagenumbering{arabic} 
\setcounter{page}{1}

The discovery of the 125 GeV Higgs boson in 2012  \cite{HiggsDiscoveryATLAS,HiggsDiscoveryCMS} has marked the starting point of a new era in particle physics, that of Higgs precision measurements, thus leading to a joint effort by both theorists and experimentalists, in order to unravel the true nature of the Higgs and its possible connection to New Physics (NP) beyond the SM (BSM). 

The ATLAS and CMS experiments have made since then outstanding progress in the measurements of the various Higgs production and decay modes, which 
serve as an important testing ground of the SM. 
Indeed, the current status is that the measured Higgs signals 
are largely compatible with the SM within the errors; in some channels the combined precision is of ${\cal O}(10\%)$ 
at the $1\sigma$ level  \cite{muVbbATLAS,muVbbCMS,muFWWATLAS,muFWWCMS,muFZZATLAS,muFZZCMS,muFgamgamATLAS,muFgamgamCMS,muFtautauATLAS,muFtautauCMS,ATLASmumu,CMSmumu}. 
In particular, in the SM the main Higgs production modes 
include the dominant gluon-fusion channel (predominantly through the top-quark loop), $gg \to h$, as well as  
the $hV$ and Vector-Boson-Fusion (VBF) channels, $q \bar q \to V \to hV$ and $VV \to h$ (the overall hard process being $qq \to qqh$), respectively. Its dominant decay mode is $h \to b \bar b$, which is currently measured only via the sub-dominant $hV$ production mode (due to the large 
QCD background in the leading model $gg \to h \to b \bar b$).
The best sensitivity, at the level of ${\cal O}(10\%)$ (combining the ATLAS and CMS measurements \cite{muFWWATLAS,muFWWCMS,muFZZATLAS,muFZZCMS,muFgamgamATLAS,muFgamgamCMS}) is currently obtained in the Higgs decay channels to vector-bosons $h \to \gamma \gamma, ZZ^\star, WW^\star$, when it is produced in the gluon-fusion channel. 

Thus, by looking for patterns of deviations in the Higgs properties, these so-called "Higgs-signals" can, therefore, shed light on the UV theory which underlies the SM. 
Indeed, the Higgs plays an important role in many of the popular BSM scenarios 
that attempt to address the fundamental shortcomings of the SM, such as the hierarchy and flavor problems, dark matter and neutrino masses. For that reason Higgs physics  
has been studied within several well motivated BSM  
scenarios such as supersymmetry (SUSY) \cite{MartinRPC1,RosiekRPC2,Decoupling}, which addresses the hierarchy problem and 
Composite Higgs-models \cite{CompositeHiggs}, in which 
the Higgs is identified as a pseudo-Nambu-Goldstone boson associated with the breaking of an underlying global symmetry. The Higgs has also been extensively studied
in a model-independent approach, using the so called SMEFT framework \cite{SMEFT}, where it is incorporated in higher dimensional operators and in Higgs-portal models
\cite{HiggsPortal1,HiggsPortal2,HiggsPortal3,HiggsPortal4,HiggsPortal5}, which address the dark matter problem.

It is widely accepted that perhaps the most appealing BSM theoretical concept is SUSY, 
since it essentially eliminates the gauge hierarchy problem 
(leaving perhaps a little hierarchy in the SUSY fundamental parameter space) and it elegantly addresses  
the unification of forces, as well as providing a well motivated dark matter candidate. Unfortunately, 
no SUSY particles have yet been observed, so that the typical SUSY scale is now pushed to the multi-TeV 
range, with the exception of some of the Electro-Weak (EW) interacting SUSY states, 
such as the lightest gauginos and SU(2) sleptons doublets. 
An interesting variation of SUSY, which is in fact a more general SUSY framework,  
includes R-parity violating (RPV) interactions \cite{RPVSUSY}. Indeed, one of the 
key incentives of the
the RPV SUSY framework is the fact that it also addresses the generation of 
neutrino masses in a distinctive manner. Many studies on RPV SUSY have, therefore, focused on reconstructing the neutrino masses and oscillation data \cite{neutrinomass1,neutrinomass2,neutrinomass3,neutrinomass3_1,neutrinomass4,neutrinomass5,neutrinomass6,neutrinomass6_1,B3_neutrinomass_constraints,neutrinomass7,neutrinomass8,neutrinomass9,neutrinomass10,neutrinomass11,neutrinomass12,neutrinomass13}, while far less attention has been devoted to the role that RPV SUSY 
may play in Higgs Physics \cite{RPVHiggsValle,RPVHiggsRosiek1,RPVHiggsParameterization,3bodyBRPVdecays1,Arhrib,BRPVdiphoton,triRPV2loopmh,BRPVDiHiggs,RPVBminusL,RPVNMSSM,UnusualRPVHiggs,VBFRPVHiggs}. 
It has also been recently proposed, that an effective RPV SUSY scenario approach involving only the
third generation \cite{3rdGenSUSY} can simultaneously explain the $R_{D^{(*)}}$ anomaly related to
B-physics and also alleviate the Hierarchy problem of the SM \cite{RPVbanomalies}. For related efforts tackling the recent B-anomalies within the RPV SUSY framework see \cite{RPVBanomaly1,RPVBanomaly2,RPVBanomaly3}.

From the experimental side, since RPV entails the decay of the LSP, the searches for RPV signatures at the LHC are based on a different strategy than in traditional SUSY channels \cite{RPVsearchesReview, RPVsearchatLHCdreiner, TRPVcharginodecays, RPVBminusLgauginodecays,RPVPhenoLSP,SameSignDileptonRPV,RPVsignaturesLHC,LongLivedSearchesRPV}. 

In this paper, we propose to interpret the observed 125 GeV Higgs-like state as a Higgs-sneutrino mixed state of the RPV SUSY framework \cite{HiggsSneutrinoMixingEilamMele,HiggsSneutrinoMixingEilamMele2} (throughout the paper we will loosely refer to the Higgs-sneutrino mixed state as the "Higgs"). We thus study the implications and effects of RPV SUSY on the 125 GeV Higgs signals. Guided by the current non-observation of SUSY particles at the LHC, 
we adopt a heavy SUSY spectrum with multi-TeV squarks and SU(2) scalar singlets as well as 
the decoupling limit in the SUSY Higgs sector. We find that, in contrast to the R-parity 
conserving (RPC) heavy SUSY scenario, where 
the lightest CP-even Higgs is SM-like, the RPV interactions 
can generate appreciable deviations from the SM rates in some production and decay channels of the lightest 125 GeV 
Higgs-sneutrino mixed state. These are generated by either Bilinear RPV (BRPV) interactions or BRPV combined with Trilinear RPV (TRPV) interactions. In particular, we find that notable effects 
ranging from ${\cal O}(20-30\%)$ up to ${\cal O}(100\%)$ may be expected in the Higgs signal strength observables in the channels, 
$p p \to h \to \mu^+ \mu^-, \tau^+ \tau^-$ and in the di-photon signal $pp \to h \to \gamma \gamma$ and  
that new sizable signals (see eq.~\eqref{eq:BRPVdec2}) 
associated with BRPV Higgs decays to gauginos, 
$h \to \tau^\pm \ell^\mp + \missET$ ($\ell =e,\mu,\tau$), may also occur in this scenario.
We study these Higgs production and decay channels under all the available constraints 
on the RPV SUSY parameter space. 

The paper is organized as follows: in section \ref{sec:The-Model} we briefly describe the RPV SUSY framework and in section \ref{sec:Higgs-signals} we layout our notation and give 
an overview of the measured signals of the 125 GeV Higgs-like state. In sections \ref{sec:Numerical-results} and \ref{sec:TRPV} we present our analysis and results for the BRPV and TRPV 
Higgs signals, respectively, and in section \ref{sec:Summary} 
we summarize. In Appendix~\ref{sec:Higgs Couplings} we give the relevant RPV Higgs 
couplings, decays and production channels, while in Appendix~\ref{sec:SUSY spectra} we list the SUSY spectra associated with the RPV SUSY benchmark models studied in the paper. 

\section{\label{sec:The-Model}The RPV SUSY framework}

The SUSY RPC superpotential is (see e.g.,  \cite{MartinRPC1,Rosiek:1989rs,RosiekRPC2,Decoupling}):
\begin{equation}
{\cal W}_{\rm RPC} = \epsilon_{ab} \left[ \frac{1}{2} h_{jk} {\hat
H}_d {\hat L}_j {\hat E}_k + h_{jk}^\prime {\hat H}_d
{\hat Q}_j {\hat D}_k + h_{jk}^{\prime \prime} {\hat H}_u
{\hat Q}_j {\hat U}_k  - \mu {\hat H}_d {\hat H}_u \right] ~,
\label{eq:lrpc}
\end{equation}

\noindent where  $\hat H_u(\hat H_d)$ are the up(down)-type Higgs
supermultiplet and ${\hat L}({\hat E}^c)$ are the leptonic SU(2)
doublet(charged singlet) supermultiplets. The ${\hat Q}$ are quark
SU(2) doublet supermultiplets and ${\hat U}^c({\hat D}^c)$ are
SU(2) up(down)-type quark singlet supermultiplets. Also, 
$j,k=1,2,3$ are generation labels and SU(2) contractions 
are not explicitly shown. 

If R-parity is violated, then both lepton and baryon numbers 
may no longer be conserved in the theory. In particular, 
when lepton number is violated then the $\hat L$
and $\hat H_d$ superfields, which have the same gauge quantum
numbers, lose their identity since there is no additional quantum
number that distinguishes between them. One can then construct
additional renormalizable RPV interactions
simply by replacing $\hat H_d \to \hat L$ in \eqref{eq:lrpc}. Thus,
the SUSY superpotential can violate lepton number (or more
generally R-parity) via  Yukawa-like
trilinear  term (TRPV) and/or a
mass-like  bilinear RPV  term (BRPV) as follows (see e.g., \cite{RPVSUSY,Roy:1992ps,Bhattacharyya:1996nj,Dreiner:1997uz,Roy:1997ca,MohapatraRPV,SarahTRPVlink}):
\begin{equation}
{\cal W}_{RPV(\lp)} \supset \frac{1}{2}
\lambda_{ijk} {\hat L}_i {\hat L}_j {\hat E}_k
 + \lambda_{ijk}^\prime {\hat L}_i {\hat Q}_j {\hat D}_k - \epsilon_i {\hat L}_i \hat H_u  ~.
\label{eq:lp}
\end{equation}
where $\lambda_{ijk}$ is anti-symmetric in the first two indices $i \neq j$ due to SU(2) gauge invariance (here also SU(2) labels are not shown).

Moreover, if R-parity is not conserved then, in addition to the
usual RPC soft SUSY breaking terms, one must also add new
trilinear and bilinear soft terms corresponding to the RPV terms
of the superpotential, e.g., to the ones in \eqref{eq:lp}. For our
purpose, the relevant ones to be added to the SUSY scalar
potential are the following soft breaking mass-like terms
\cite{neutrinomass6,neutrinomass4,neutrinomass1,Roy:1996bua,Diaz:1997xc,Chang:1999ih,Mukhopadhyaya:1999gy,Davidson:1999mc}:

\begin{equation}
V_{BRPV} = (M_{LH}^2)_i \tilde L_i H_d -
(B_\epsilon)_i {\tilde L}_i H_u \label{eq:bterm}~,
\end{equation}

\noindent where $\tilde L$ and $H_d$ are the scalar components of
$\hat L$ and $\hat H_d$, respectively.

In what follows, we will consider a single generation 
BRPV scenario, i.e., that in eq.~\eqref{eq:bterm} R-Parity is violated only among the interactions of a single slepton. 
In particular, we will be focusing  
mainly on the 3rd generation bilinear soft breaking mixing term, $(B_\epsilon)_3$, which mixes the 3rd generation (tau-flavored) left-handed slepton neutral and charged fields with the neutral and charged up-type Higgs fields, respectively.
The bilinear soft term $(B_\epsilon)_3$ leads in general to a non-vanishing VEV of the tau-sneutrino, $\left< \tilde\nu_\tau \right> = v_{\tilde\nu_\tau}$. However,
since lepton number is not a conserved quantum number in this scenario,the $\hat H_d$ and $\hat L_3$ superfields lose their identity
and can be rotated to a particular basis $(\hat H_d^\prime, \hat L_3^\prime)$ in which either $\epsilon_3$ or $v_3 =v_{\tilde\nu_\tau}$ are set to zero \cite{neutrinomass1,Davidson:1999mc,Ferrandis:1998ii,Davidson:2000ne,Grossman:2000ex}. 
In what follows,
we choose for convenience to work in the “no VEV” basis, 
$v_{\tilde\nu_\tau}=0$, which simplifies our analysis below.
In this basis the minima conditions in the scalar potential read (we follow below the notation 
of the package \texttt{SARAH} \cite{Sarah,Sarah1,Sarah2} and use some  
of the expressions given in 
\cite{Sarahlink}): 
\begin{eqnarray}
&& 1) ~~ m_{H_{d}}^{2}v_{d}-v_{u}B_{\mu}+\frac{1}{8}\left(g_{1}^{2}+g_{2}^{2}\right)v_{d}\left(-v_{u}^{2}+v_{d}^{2}\right)+\left|\mu\right|^{2}v_{d} =0 ~, \label{eq:minimaconditionI} \\
&& 2) ~~ -\frac{1}{8}\left(g_{1}^{2}+g_{2}^{2}\right)v_{u}\left(-v_{u}^{2}+v_{d}^{2}\right)+\frac{1}{2}\left(-2v_{d}B_{\mu}+2v_{u}\left(m_{H_{u}}^{2}+\left|\mu\right|^{2}+\left|\epsilon_3\right|^{2}\right)\right)  =0 ~, \label{eq:minimaconditionII}\\
&& 3) ~~ \left( m_{LH}^{2} \right)_3 + \left(B_{\epsilon} \right)_3 \tan\beta-\epsilon_3\mu =0 ~, \label{eq:minimaconditionIII}
\end{eqnarray}
where $B_\mu$ is the soft breaking bilinear term $B_\mu H_d H_u$ (i.e., corresponding to the $\mu$-term in the superpotential) and $v_u,v_d$ are the VEV's of the up and down Higgs fields, $H_u^0,H_d^0$. 

Without loss of generality, in what follows, we parameterize $B_{\mu}$ in terms of the physical pseudo-scalar mass 
$m_{A}$ using the RPC MSSM relation
$m_{A}^{2} \equiv\csc\beta\sec\beta\, B_{\mu}$ (see e.g., \cite{RPVHiggsParameterization}), thus
defining the soft BRPV "measure" as (from now on and 
throughout the rest of the paper we drop the generation index of the BRPV terms): 
\begin{align}
\delta_B \equiv \frac{B_\epsilon}{B_\mu} ~,
\end{align}
so that $B_{\epsilon}$ will be given in terms of a new BRPV parameter $\delta_{B}$: 

\begin{align}
B_{\epsilon} & \equiv\delta_{B}B_{\mu}=
\frac{1}{2} m_{A}^{2} \sin\left(2\beta\right) \delta_{B} ~. \label{eq:B_redefinition}\end{align}

We also define, in a similar way, the measure of BRPV in the superpotential, $\delta_{\epsilon}$, via:
\begin{align}
\epsilon & 
\equiv\delta_{\epsilon}\mu ~.
\label{eq:eps_redefinition}
\end{align}

\subsection{\label{sub:snuhiggs}The RPV SUSY scalar sector and Higgs-Sneutrino mixing}

Using eqs.~\eqref{eq:minimaconditionI}--\eqref{eq:B_redefinition},
the induced CP-odd and CP-even scalar mass matrices squared reads (see e.g., \cite{Sarahlink}):\footnote{The CP-odd scalar mass matrix, $m^2_O$, has a massless state which corresponds to the Goldstone boson.}$^{,}$\footnote{We note that 
too large 
values of $\delta_{B}$ may drive (depending on the other free-parameters in the Higgs sector) the lightest 
mass-squared eigenstates of both the CP-even and CP-odd mass matrices to non-physical negative values. We will thus 
demand non-negative mass-squared physical 
eigenvalues 
for the CP-even and CP-odd physical states by 
bounding $\delta_{B}$ accordingly.}
\begin{align}
m_{O}^{2} & =\left(\begin{array}{ccc}
s_{\beta}^{2}m_{A}^{2} & m_{A}^{2}s_{\beta}c_{\beta} & -\delta_{B}m_{A}^{2}s_{\beta}^{2}\\
m_{A}^{2}s_{\beta}c_{\beta} & c_{\beta}^{2}m_{A}^{2} & -\delta_{B}m_{A}^{2}s_{\beta}c_{\beta}\\
-\delta_{B}m_{A}^{2}s_{\beta}^{2} & -\delta_{B}m_{A}^{2}s_{\beta}c_{\beta} & m_{\tilde{\nu}_{\tau}}^{2}\end{array}\right)
\label{eq:CP-oddscalarmassmatrix}
\end{align}
\begin{align}
m_{E}^{2} & =\left(\begin{array}{ccc}
s_{\beta}^{2}m_{A}^{2}+m_{Z}^{2}c_{\beta}^{2}+\delta_{11}^{t-\tilde{t}} & -s_{\beta}c_{\beta}m_{A}^{2}-m_{Z}^{2}s_{\beta}c_{\beta}+\delta_{12}^{t-\tilde{t}} & -\delta_{B}m_{A}^{2}s_{\beta}^{2}\\
-s_{\beta}c_{\beta}m_{A}^{2}-m_{Z}^{2}s_{\beta}c_{\beta}+\delta_{12}^{t-\tilde{t}} & c_{\beta}^{2}m_{A}^{2}+m_{Z}^{2}s_{\beta}^{2}+\delta_{22}^{t-\tilde{t}} & \delta_{B}m_{A}^{2}s_{2\beta}/2\\
-\delta_{B}m_{A}^{2}s_{\beta}^{2} & \delta_{B}m_{A}^{2}s_{2\beta}/2 & m_{\tilde{\nu}_{\tau}}^{2}\end{array}\right)\label{eq:CP-even perturbed mass matrix}
\end{align}
with $s_{\beta}\equiv\sin\beta$, $c_{\beta}\equiv\cos\beta$ and $\tan\beta = v_u/v_d$. The $\tau$-sneutrino mass term, $m_{\tilde{\nu}_{\tau}}$, is given in the RPV SUSY framework by: 
\begin{align}
m_{\tilde{\nu}_{\tau}}^{2} & =\underbrace{m_{\tilde\tau_{LL}}^{2}+\frac{1}{8}\left(g_{1}^{2}+g_{2}^{2}\right)\left(-v_{u}^{2}+v_{d}^{2}\right)}_{\left(m_{\tilde{\nu}_{\tau}}^{2}\right)_{RPC}}+
\left|\epsilon\right|^{2}\label{eq:msndependencies} ~,
\end{align}
where $m_{\tilde\tau_{LL}}$ is the soft 
left-handed slepton mass term, $m_{\tilde\tau_{LL}}^2 \tilde L_3^\star \tilde L_3$. We have denoted in eq.~\eqref{eq:msndependencies} 
the $\tau$-sneutrino mass term in the RPC limit $\epsilon \to 0$ by ${\left(m_{\tilde{\nu}_{\tau}}^{2}\right)_{RPC}}$ 
(note that the correction to $m_{\tilde{\nu}_{\tau}}^{2}$ in the BRPV framework is $(\Delta m_{\tilde{\nu}_{\tau}}^{2})_{RPV} = |\epsilon|^2$). 
Note also that we 
have added in the CP-even sector the dominant top-stop loop
corrections, $\delta_{ij}^{t-\tilde{t}}$, which are required in order to lift the lightest Higgs
mass to its currently measured value \cite{FeynHiggs};
see also discussion on the Higgs mass filter in section \ref{sec:Numerical-results}. 

The physical CP-even mass eigenstates in the RPV framework, which we denote below by $S^E_{RPV} = \left(h_{RPV},\, H_{RPV},\,\tilde{\nu}_{RPV}\right)^T$, are obtained upon diagonalizing the CP-even scalar mass-squared matrix:

\begin{align}
S^E = Z^E S^E_{RPV} ~,
\label{eq:snRPCweakstate}
\end{align}

where
$S^E = \left(H_{d},\, H_{u\,},\,\tilde{\nu}_{\tau}\right)^T$ are the corresponding weak states of in the CP-even Higgs sector and $Z_E$ is 
the unitary $3 \times 3$ matrix which diagonlizes
$m_{E}^{2}$, defined here as:
\begin{align}
Z^E & =\left(\begin{array}{ccc}
    Z_{h1}   & Z_{H1} & Z_{\tilde\nu 1}  \\
    Z_{h2}   & Z_{H2} & Z_{\tilde\nu 2}  \\
    Z_{h3}   & Z_{H3} & Z_{\tilde\nu 3}  \end{array}\right) ~.
\label{eq:ZE}
\end{align}

We would like to emphasize a few aspects and features of the BRPV SUSY framework which are manifest in the CP-even Higgs mass matrix and spectrum and are of considerable importance for our study in this paper: 
\begin{itemize}
\item We will be interested in
the properties (i.e., production and decay modes) of $h_{RPV}$, which is the lightest CP-even Higgs state in the BRPV framework. This state has a Sneutrino component due to the 
Higgs-sneutrino mixing terms (i.e., $\propto \delta_B$) in the CP-even Higgs mass matrix \cite{HiggsSneutrinoMixingEilamMele,HiggsSneutrinoMixingEilamMele2}. 
The element $Z_{h3}$ is the one which corresponds to the Sneutrino component in $h_{RPV}$ and is, therefore, responsible for the $\tilde{\nu}_{\tau} - h$ mixing 
phenomena.  It 
depends on $\delta_{B}$ 
and thus shifts some of the RPC light-Higgs couplings, as will be discussed below. 
In particular, we interpret the observed 125 GeV 
Higgs-like state as the lightest Higgs-sneutrino mixed state  
$h_{RPV}$ and, in our
numerical simulations below, we demand 
$122<m_{h_{RPV}}<128$ GeV, in accordance with the LHC data where we allow some room for 
other SUSY contributions to the Higgs mass, i.e., beyond the simplified RPV framework discussed in this work.
\item The elements $Z_{h1}$ and $Z_{h2}$ correspond to the $H_{d}^0$ and $H_{u}^0$ components in $h_{RPV}$. They are independent of the soft BRPV parameter $\delta_{B}$ at $\mathcal{O}\left(\delta_{B}\right)$, so that, at leading order in $\delta_{B}$, they are the same as the corresponding RPC elements.  
\item Guided by the current non-observation of new sub-TeV heavy Higgs states at the LHC, we will assume  
the decoupling limit in the SUSY Higgs sector \cite{DecouplingHaber,DecouplingCarena,Decoupling}, 
in which case the RPC Higgs couplings are SM-like. 
We will demonstrate below that the BRPV effects may be 
better disentangled in this case.  
\end{itemize}

\subsection{\label{sub:Neutralino-mass-matrix}The Gaugino sector}

With the BRPV term in the superpotential ($\epsilon \hat L \hat H_u$ in eq.~\eqref{eq:lp})  
and assuming only 3rd generation BRPV, i.e., only $\epsilon_3 \neq 0$, the neutralinos and charginos mass matrices read:  
\begin{align}
m_{N} & =\left(\begin{array}{cc}
\left(m_{\nu_{\tau}}\right)_{loop}^{\delta_{B}\delta_{B}}+\left(m_{\nu_{\tau}}\right)_{loop}^{\delta_{B}\delta_{\epsilon}} & V_N^{BRPV}\\
(V_N^{BRPV})^T & m_N^{RPC}  \end{array}\right)
\label{eq:Neutralinomassmatrix} ~,
\end{align}
\begin{align}
m_{C} & =\left(\begin{array}{cc}
m_{\tau} & V_C^{BRPV} \\
\vec{0}^T & m_C^{RPC} \end{array}\right)
\label{eq:charginomassmatrix} ~,
\end{align}
where $V_N^{BRPV} \equiv (0,0,0,\delta_{\epsilon}\mu)$, 
$V_C^{BRPV} \equiv (0,-\delta_{\epsilon}\mu)$,
$\vec{0}=(0,0)$ and $\delta_\epsilon = \epsilon/\mu$ (dropping the generation index, see eq.~\eqref{eq:eps_redefinition}). Also, $(m_{C})_{11} = m_\tau$ is the bare mass of
the $\tau$-lepton and in $(m_{N})_{11}$ we have added the loop-induced
BRPV contributions $\left(m_{\nu_{\tau}}\right)_{loop}^{\delta_{B}\delta_{B}}$
and $\left(m_{\nu_{\tau}}\right)_{loop}^{\delta_{B}\delta_{\epsilon}}$
to the tau-neutrino mass \cite{B3_neutrinomass_constraints}
(which is used 
in section~\ref{sec:Numerical-results} 
in order to constrain the BRPV parameters $\delta_{B}$ and $\delta_{\epsilon}$). 
Finally, $m_N^{RPC}=m_N^{RPC}(M_1,M_2,\mu,m_Z,\tan\beta,s_W)$ 
and $m_C^{RPC}=m_C^{RPC}(M_2,\mu,m_Z,\tan\beta,s_W)$ 
are the $4 \times 4$ neutralino mass matrix 
and the $2 \times 2$ chargino mass matrix in the RPC limit, respectively,   
which depend on the U(1) and SU(2) gaugino mass terms $M_1$ and $M_2$, 
on the bilinear RPC $\mu$ term, on $\tan\beta$ and on the Z-boson mass $m_Z$ and the Weinberg angle $\theta_W$ (see e.g., \cite{Sarahlink}). 

The physical neutralino and chargino states, 
$F^{\tilde\chi^0}_{RPV}=(\tilde\chi^0_1,\tilde\chi^0_2,\tilde\chi^0_3,\tilde\chi^0_4,\tilde\chi^0_5)^T$ and 
$F^{\chi^\pm}_{RPV}=(\chi^\pm_1,\chi^\pm_2,\chi^\pm_3)^T$, respectively,  
are  
obtained by diagonalizing their mass matrices in \eqref{eq:Neutralinomassmatrix} and \eqref{eq:charginomassmatrix}. For the neutralinos we have (i.e., with only 3rd generation neutrino-neutralino mixing):    
%
\begin{align}
F^{\tilde\chi^0} = U_N F^{\tilde\chi^0}_{RPV} ~,
\label{eq:UN}
\end{align}
%
where $F^{\tilde\chi^0} = \left(\nu_{\tau},\tilde{B},\tilde{W},\tilde{H}_{d},\tilde{H}_{u}\right)^T$ are the neutralino weak states 
and $U_{N}$ is the unitary matrix which diagonlizes
$m_{N}$ in \eqref{eq:Neutralinomassmatrix}. In particular, 
we identify the lightest neutralino state 
in the RPV setup, $\tilde\chi^0_1$, as
the $\tau$-neutrino $\tilde\chi^0_1 \equiv \nu_{\tau}$. 
Note that the entries $(U_N)_{ij}$ enter in the
Higgs couplings to a pair of neutralinos and, in particular,
generates the coupling $h \nu_\tau \tilde{\chi}_{2}^0$ ($h \equiv h_{RPV}$), where 
$\tilde\chi_2^0$ is the 2nd lightest neutralino state corresponding to the lightest neutralino in the RPC case
(see Appendix~\ref{subsec:Higgs couplings and decays to gauginos}). As will be discussed in section \ref{sec:Numerical-results}, this new RPV coupling
opens a new Higgs 
decay channel $h \rightarrow \nu_\tau \tilde{\chi}_2^0$, if
$m_{\tilde\chi_2^0} < m_h$ and also 
enters in the the loop-induced contribution to $m_{\nu_{\tau}}$.

In the chargino's case, 
since the matrix $m_{C}$ is not symmetric, it is diagnolized with the
singular value decomposition procedure, 
which ensures a positive mass spectrum:
\begin{align}
U_{L} m_{C} U_{R}^{\dagger} & =m_{C}^{diag} ~. \label{eq:SVD_chargino}
\end{align}

The chargino physical states $\left(\chi_{i}^{\pm}\right)$ are then obtained 
from the 
weak states $F^{\chi^-} = \left(\tau_{L},\tilde{W}^{-},\tilde{H}_{d}^{-}\right)^T$ and $F^{\chi^+}= \left(\tau_{R},\tilde{W}^{+},\tilde{H}_{u}^{+}\right)^+$ by:
%
\begin{align}
F^{\chi^-} = (U_L)^T F^{\chi^-}_{RPV} ~~,~~
F^{\chi^+} = U_R F^{\chi^+}_{RPV}
\label{eq:ULUR} ~,
\end{align}
%
%
where, here also, the lightest chargino is identified as the $\tau$-lepton, i.e., $\tau^+ = \chi^{+}_{1}$ and 
the elements of the chargino rotation matrices $U_{L,R}$ enter in the Higgs couplings to a pair of charginos. Thus, the decay $h \rightarrow\chi_{1}^{+} \chi_{1}^{-}$ corresponds in the RPV framework to 
$h\rightarrow\tau^{+}\tau^{-}$. In addition, if
$m_{\tau}+m_{\chi_{2}^{+}} < m_{h}$, then the decay 
$h \rightarrow \tau^\pm \chi_{2}^{\mp}$ (i.e., 
the decay $h\rightarrow\chi_{1}^{\pm} \chi_{2}^{\mp}$)
is also kinematically open (see Appendix~\ref{subsec:Higgs couplings and decays to gauginos}). 

\section{\label{sec:Higgs-signals}The 125 GeV Higgs signals}

The measured signals of the 125 GeV Higgs-like particle are sensitive to a variety of new physics scenarios, which may alter the Higgs couplings to the known SM particles that are involved in its production and decay channels. 

We will use below the
Higgs ``signal strength" parameters, which are defined as the ratio
between the Higgs production and decay rates and their SM expectations:
\begin{eqnarray}
\mu_{if}^{(P)} \equiv
\mu_{i}^{(P)} \cdot \mu_{f} \cdot \frac{\Gamma^{h}_{SM}}{\Gamma^{h}}
 ~, \label{eq:HiggsObservablemuij}
\end{eqnarray}
where $\mu_{i}^{(P)}$ and $\mu_{f}$ are the normalized production and decays factors which, in the narrow Higgs width approximation, read:
\begin{eqnarray}
\mu_i^{(P)}  = \frac{\sigma(i \to h)}{\sigma(i \to h)_{SM}} ~~,~~ 
\mu_f = \frac{\Gamma(h \to f)}{\Gamma(h \to f)_{SM}} ~,\label{eq:Dfac}
\end{eqnarray}
and $\Gamma^h$ $(\Gamma^h_{SM})$ is the total
width of the 125 GeV Higgs (SM Higgs). Also, 
$i$ represents the parton content in the proton
which is
involved in the production mechanism and $f$ is the Higgs
decay final state.

We will consider below the signal strength signals 
which are associated with the leading hard production mechanisms: gluon-fusion, $gg \to h$, $hV$ production, $q \bar q \to V \to hV$, and VBF, 
$VV \to h$.\footnote{We neglect Higgs production
via $pp \to t \bar t h$, which,
although included in the ATLAS and CMS fits,
is 2-3 orders of magnitudes smaller than the gluon-fusion channel. Note that additional sources of Higgs production via heavy SUSY scalar decays may be present as well \cite{SoniHeavyScalarpaper}.} The $q \bar q$-fusion 
production channel, which is negligible in the SM due to the vanishingly small SM Yukawa couplings of the light-quarks, will be considered for the TRPV scenario in the next section.
We will use the usual convention, denoting by $i=F$ 
the gluon-fusion channel and by $i=V$ the $hV$ and VBF 
channels; for clarity and consistency with the above definitions, we will also explicitly denote the underlying 
hard production mechanism by a bracketed superscript. 
The decay channels that will be considered below are 
$h \to \gamma \gamma,~W W^\star,~Z Z^\star$ and 
$h \to \mu^+ \mu^-,\tau^+ \tau^-,~b \bar b$. 

In particular, in the BRPV SUSY scenario we have:
\begin{eqnarray}
\mu_F^{(gg)} &=& \frac{\Gamma(h \to gg)}{\Gamma(h \to gg)_{SM}} ~, \label{eq:ggfuse} \\
\mu_V^{(hV)} &=& \mu_V^{(VBF)} = \left( g_{hVV}^{RPC} \right)^2 ~, \label{eq:prodfac}
\end{eqnarray}
for the production factors and
\begin{eqnarray}
\mu_{bb} &=& \left( g_{hbb}^{RPC} \right)^2 ~, \\
\mu_{VV^\star} &=& \left( g_{hVV}^{RPC} \right)^2 ~, \\
\mu_{\mu \mu / \tau \tau} &=& \frac{\Gamma(h \to \mu^+ \mu^- / \tau^+ \tau^-)}{\Gamma(h \to \mu^+ \mu^- / \tau^+ \tau^-)_{SM}} ~ \\
\mu_{\gamma \gamma} &=& \frac{\Gamma(h \to \gamma \gamma)}{\Gamma(h \to \gamma \gamma)_{SM}}  ~, \label{eq:decfac}
\end{eqnarray}
for the decay factors, where $VV^\star = W W^\star,~Z Z^\star$ and the RPC $hVV$ and $hbb$ couplings, $g_{hVV}^{RPC}$ 
and $g_{hbb}^{RPC}$, as well as the decay widths for 
$h \to gg, \gamma \gamma, \mu^+ \mu^-, \tau^+ \tau^-$ are given in Appendix \ref{sec:Higgs Couplings}. 
In particular, 
the $hV$ and VBF production channels as well as 
the Higgs decays to a pair of $W$ and $Z$ bosons are not changed in our BRPV setup (i.e., in the no-VEV basis $\left< \tilde\nu_\tau \right> =0$) with respect to the RPC SUSY framework. 
For the total Higgs width in the RPV SUSY scenario we add the new decay channels  
$h \to \tau^\pm \chi_2^\mp$ and $h \to \nu_\tau \tilde\chi_2^0$ 
when they are kinematically open (see next section). 

Finally, in the numerical simulations presented below we use the combined ATLAS and CMS signal strength measurements 
(at 13 TeV) which are listed in Table~\ref{tab:Higgs-filters}.

\begin{table}
    \caption{\label{tab:Higgs-filters}
    Combined ATLAS and CMS (13 TeV) signal strength measurements corresponding to the Higgs observables defined in eq.~\eqref{eq:HiggsObservablemuij}. We have closely followed the Higgs data as summarized in Table I in \cite{HiggsSignalsfilters}, with some updated more recent results where needed (citations to the relevant papers are given in the 3rd column). Note that we have added the recent combined signal strength measurement 
    in the $pp \to h \to \mu \mu$ channel,  $\mu_{F\mu\mu}^{(gg)}$. Also, in each channel, we have indicated (with a superscript) the specific hard production channel ($gg$, $hV$ or VBF), see also text.}
   \medskip{}
    \centering{}
    \begin{tabular}{lcl}
        \toprule 
        
        & \llap{A}TLAS + CM\rlap{S} &  \tabularnewline
        \midrule
        \addlinespace
        $\mu_{Vbb}^{(hV/hW)}$ & $1.07_{-0.22}^{+0.23}$ & \cite{muVbbATLAS,muVbbCMS}   \tabularnewline
        \addlinespace
        $\mu_{Vbb}^{(hZ)}$ & $1.20_{-0.31}^{+0.33}$ & \cite{muVbbATLAS}   \tabularnewline
        \addlinespace
        $\mu_{FWW}^{(gg)}$ & $1.24_{-0.16}^{+0.15}$ & \cite{muFWWATLAS,muFWWCMS}   \tabularnewline
        \addlinespace
        $\mu_{FZZ}^{(gg)}$ & $1.09^{+0.11}_{-0.11}$ & \cite{muFZZATLAS,muFZZCMS}   \tabularnewline
        \addlinespace
        $\mu_{F\gamma\gamma}^{(gg)}$ & $1.02_{-0.11}^{+0.12}$ & \cite{muFgamgamATLAS,muFgamgamCMS}  \tabularnewline
        \addlinespace
        $\mu_{F\tau\tau}^{(gg)}$ & $1.06^{+0.40}_{-0.37}$ & \cite{muFtautauATLAS,muFtautauCMS}  \tabularnewline
        \addlinespace
        $\mu_{V\gamma\gamma}^{(VBF)}$ & $1.10_{-0.31}^{+0.36}$ & \cite{muFgamgamATLAS,muFgamgamCMS}  \tabularnewline
        \addlinespace
        $\mu_{V\tau\tau}^{(VBF)}$ & $1.15^{+0.36}_{-0.34}$ & \cite{muFtautauATLAS,muFtautauCMS}  \tabularnewline
        \addlinespace
        $\mu_{F\mu\mu}^{(gg)}$ & $0.55_{-0.70}^{+0.70}$ & \cite{ATLASmumu,CMSmumu}  \tabularnewline
        \bottomrule
     
        \addlinespace
    \end{tabular}
\end{table}

\section{\label{sec:Numerical-results}Bilinear RPV - Numerical results}

To quantify the impact of BRPV on the 125 GeV Higgs physics 
we performed a numerical simulation, evaluating all relevant Higgs production and decays modes under the following 
numerical and parametric setup (for recent work in this spirit see \cite{BRPVdiphoton,triRPV2loopmh}):
\begin{itemize}
    \item Our relevant input parameters are $\bigl(\mu, M_{1}, M_{2}, t_{\beta}, m_{A}, m_{\tilde{\nu}_{\tau}},$ $m_{\tilde{q}}, \tilde{A}, m_{\tilde{b}_{RR}}, m_{\tilde{\tau}_{RR}}, \delta_{\epsilon}, \delta_{B} \bigr)$, where $m_{\tilde{q}}$ is used as a common  left-handed (soft) squark mass (i.e., $m_{\tilde{q}} \tilde{q}_L^\star \tilde{q}_L$) for both the stop and sbottom states (see Appendix~\ref{subsec:Higgs couplings to squarks and sleptons}) and $t_{\beta} \equiv \tan\beta $. 
    \item In the stop sector we have assumed a degeneracy between the right and left-handed soft masses, i.e.,  $m_{\tilde{t}_{RR}}=m_{\tilde{q}}$; this is also used 
    in the calculation of the stop-top loop corrections to the CP-even scalar mass matrix, see~\cite{FeynHiggs}. 
    On the other hand, in the sbottom and stau sectors  we keep 
    the right-handed soft mass terms, $m_{\tilde{b}_{RR}},m_{\tilde{\tau}_{RR}}$,  
    as free-parameters.
    \item  We adopt the Minimal Flavor Violation (MFV) setup  for the squarks and sleptons soft trilinear terms, assuming that they are proportional to the corresponding Yukawa couplings: $A_{f}=y_{f} \cdot \tilde{A}$, for $f=t,b,\tau$.\footnote{For the fermion Yukawa couplings we have $y_{f}=\frac{\sqrt{2}m_{f}}{vc_{\beta}}$
for the down-type quarks and leptons and 
$y_{f}=\frac{\sqrt{2}m_{f}}{vs_{\beta}}$
for up-type quarks.} 
    We thus vary the common trilinear soft term $\tilde A$ for all the squarks and sleptons states. 
    \item We randomly vary the model input parameters 
    $\bigl(\mu, M_{1}, M_{2}, t_{\beta}, m_{A}, m_{\tilde{\nu}_{\tau}},$ $m_{\tilde{q}}, \tilde{A}, m_{\tilde{b}_{RR}}, m_{\tilde{\tau}_{RR}}, \delta_{\epsilon}, \delta_{B} \bigr)$ within fixed ranges which are listed in Table~\ref{tab:Input-parameter-ranges}. In some instances and depending on the RPV scenario analyzed below, these ranges are refined for the purpose of optimizing the BRPV 
    effect, thereby focusing on more specified regions of the RPV SUSY parameter space.
    \item In cases where the Higgs decays to gauginos, we consider light gaugino states with a mass $\sim 90-100$ GeV (see e.g. \cite{Gambit}), which requires a higgsino mass parameter $\mu$ and/or gaugino mass parameters $M_{1,2}$ of $\mathcal{O}(100 ~{\rm GeV})$.
         
    \item We impose the following set of "filters" and constraints to ensure viable model configurations: 
\begin{description}
    \item[\underline{\bf Higgs mass}]
    We fix the lightest Higgs mass to its observed value $m_{h}^{obs}\backsimeq125$ GeV in the 
    computation of the Higgs production and decay rates. 
    Nonetheless, we allow for a theoretical uncertainty of $\pm3$ GeV in the calculated Higgs mass 
    (leaving some room for other possible SUSY contributions that are not accounted for in our minimal RPV SUSY framework), thus requiring that $122\,\text{GeV}<m_{h}^{calc}<128$ GeV. In particular, we include in $m_h^{calc}$ the leading top-stop corrections (see Eq.~\eqref{eq:CP-even perturbed mass matrix}) and the sbottom and stau 1-loop contributions (which are not explicitly added in Eq.~\eqref{eq:CP-even perturbed mass matrix} but can be relevant for large $\tan\beta$ \cite{MSSMCornered}):
\begin{align}
\left(\Delta m_{h}^{2}\right)_{\tilde{f}} & \approx-\frac{N_{c}^{\tilde{f}}}{\sqrt{2}G_{F}}\frac{y_{f}}{96\pi^{2}}\frac{\mu^{4}}{m_{\tilde{f}}^{2}}\label{eq:sfermionHiggsmasscorrection} ~,
\end{align}
where here $\tilde{f}=\tilde{b},\tilde{\tau}$ , $N_{c}^{\tilde{b}}=3$,
$N_{c}^{\tilde{\tau}}=1$, $m_{\tilde{b}}^{2}=m_{\tilde{b}_{1}} \cdot m_{\tilde{b}_{2}}$
and $m_{\tilde{\tau}}^{2}=m_{\tilde{\tau}_{2}} \cdot m_{\tilde{\tau}_{3}}$
and it is understood that $m_{\tilde{\tau_{2}}}$ and $m_{\tilde{\tau_{3}}}$
are the masses of the two lightest slepton states ($\tilde{\tau}_1$ being the massless
Goldstone boson). 
%
    \item[\underline{\bf Neutrino masses}]
    The RPV parameters are subject to constraints from various processes \cite{TRPVbounds}, 
    such as flavor violating $b$-decays $b\rightarrow s\gamma$ \cite{btosg1,btosg2,btosg3,btosg4} and  
Higgs decays \cite{3bodyBRPVdecays1}, as well as radiative leptonic decays, e.g., $\mu\rightarrow e\gamma$ \cite{mutoeg1,mutoeg2}. Other notable quantities that are sensitive to the RPV parameter-space constraints are, e.g.,   
Electric Dipole Moments (EDM's) \cite{EDM1,EDM2,EDM3,Nodoka1,Nodoka3,Nodoka2} and neutrino masses \cite{neutrinomass1,neutrinomass2,neutrinomass3,neutrinomass4,neutrinomass5,neutrinomass6,B3_neutrinomass_constraints}. A recent paper reviewing the various constraints on the RPV parameter-space
is given in \cite{Arhrib} and bounds on the TRPV couplings
can be found in \cite{RPVsearchatLHCdreiner,lambdaprimeconstraints2}.

We find that the strongest constraints on the BRPV parameters $\delta_B$ and $\delta_\epsilon$ are from neutrino masses. 
In particular, neutrino masses can be generated at tree-level when only $\delta_\epsilon \neq 0$ and at the 1-loop level if also $\delta_B \neq 0$. In the former case $m_\nu \propto \delta_\epsilon^2$, while at 1-loop $m_\nu \propto \delta_B \delta_\epsilon, \delta_B^2$, see \cite{B3_neutrinomass_constraints}.     
For example, the $\delta_B^2$ 1-loop contribution to the neutrino masses, which enters in eq.~\eqref{eq:Neutralinomassmatrix} is
\cite{B3_neutrinomass_constraints} (for the expression of $\left(m_{\nu_{\tau}}\right)_{loop}^{\delta_{B}\delta_{\epsilon}}$
which is rather lengthy we refer the reader to \cite{B3_neutrinomass_constraints}):  
\begin{align}
\left(m_{\nu_{\tau}}\right)_{loop}^{\delta_{B}\delta_{B}} & =\sum_{\alpha=1}^{4}\frac{\left(\frac{\delta_{B}m_{A}^{2}s_{2\beta}}{2}\right)^{2}}{4c_{\beta}^{2}}\left(g_{2}U_{\alpha2}^{RPC}-g_{1}U_{\alpha1}^{RPC}\right)^{2}m_{\tilde{\chi}_{\alpha}} \nonumber \\
& \times \Bigl[I_{4}\left(m_{h},m_{\tilde{\nu}_{\tau}},m_{\tilde{\nu}_{\tau}},m_{\tilde{\chi}_{\alpha}}\right)\left(1-\left(c_{\beta}Z_{h1}+s_{\beta}Z_{h2}\right)^{2}\right)\nonumber \\
 & \quad +I_{4}\left(m_{H},m_{\tilde{\nu}_{\tau}},m_{\tilde{\nu}_{\tau}},m_{\tilde{\chi}_{\alpha}}\right)\left(c_{\beta}Z_{h1}+s_{\beta}Z_{h2}\right)^{2}-I_{4}\left(m_{A},m_{\tilde{\nu}_{\tau}},m_{\tilde{\nu}_{\tau}},m_{\tilde{\chi}_{\alpha}}\right)\Bigr]
\label{eq:neutrinomassatdecoupling}
\end{align}
where $U^{RPC}$ is the neutralino mixing matrix in the RPC limit (i.e., corresponding
to $m_N^{RPC}$ which is the 4X4 RPC block in \eqref{eq:Neutralinomassmatrix}), $m_{\chi_{\alpha}}$
are the neutralino masses in the RPC case, i.e., $\alpha=1,2,3,4$, and $I_{4}$ is defined in \cite{B3_neutrinomass_constraints}. Also, we
have used our definition for the BRPV parameter $\delta_B$ in eq.~\eqref{eq:B_redefinition} and $s_{\alpha-\beta}^{2}=\left(c_{\beta}Z_{h1}+s_{\beta}Z_{h2}\right)^{2}$. Furthermore, 
$m_{H}$ is the mass of the heavy CP-even Higgs state and $m_{\tilde{\nu}_{\tau}}$ is
the sneutrino mass. 

We use below the current Laboratory bounds on the 
muon and $\tau$-neutrino masses: $m_{\nu_{\mu}}< 0.19$ MeV
and $m_{\nu_{\tau}}<18.2$ MeV \cite{PDG2017}.   
In particular, in our numerical simulations, we evaluate the contribution of $\delta_{B}$ and $\delta_{\epsilon}$
to the relevant neutrino mass for each run, i.e., calculating $\left(m_{\nu_{\tau}}\right)_{loop}^{\delta_{B}\delta_{B}}$
and $\left(m_{\nu_{\tau}}\right)_{loop}^{\delta_{B}\delta_{\epsilon}}$ and requiring  
the lightest physical neutralino state ($\tilde\chi^0_1=\nu_\mu$ or $\tilde\chi^0_1=\nu_\tau$ depending on the RPV scenario considered, see eq.~\eqref{eq:Neutralinomassmatrix}) 
to have a mass below these bounds.

We note that neutrino oscillation data and cosmology, imply tighter
indirect bounds on neutrino masses; at the sub-eV range \cite{PDGcosmo}. These
bounds are, however, model dependent and they apply when interpreted
within the SM of particle physics and the standard cosmological 
framework.
In particular, oscillation data constrain the differences between the 
square
of the neutrino masses,
$\Delta m_{ij}^2 = m_{\nu_i}^2 - m_{\nu_j}^2$, and therefore it does not
exclude scenarios with e.g., degenerate MeV-scale neutrinos.
If indeed the neutrino mass scale is at the MeV range, for example, then
a fine-tuning of the order of $\mathcal{O}(10^{-12})$ in the neutrino mass
squared differences is required in order
to accommodate the neutrino oscillation data measurements or,
alternatively, some underlying flavor symmetry within a model
may be responsible for $\Delta m_{ij}^2 \ll m_{i,j}^2$.
It should be noted, though, that even without a flavor model,
a fine tuning of $\mathcal{O}(10^{-12})$ should not be dismissed a priori
since comparable and even more severe fine-tuning is
currently observed in nature, e.g., in the fermion mass spectrum,
in the gauge-Higgs sector and in the cosmological
constant. In that respect we note that
the purpose of this paper is not to reconstruct
the neutrino oscillation data within a given flavor model
but rather to study the impact of RPV SUSY on the Higgs signals
under an unbiased and model independent manner using
direct constraints on the RPV parameters.

The cosmological bounds (from Big Bang Nucleosynthesis,
the power spectrum of the Cosmic Microwave Background
anisotropies and the large scale clustering of cosmological structures),
assume the minimal $\Lambda$CDM cosmological model and
the standard neutrino decoupling process, i.e., involving only
weak interactions,
so that the only massless or light (sub-keV) relic particles
since the Big Bang Nucleosynthesis (BBN) epoch are assumed to be
photons and stable active neutrinos.
Thus, the cosmology bounds do not apply within extended scenarios
involving extra light particles (relics) or
unstable neutrinos with a relative short life-time \cite{CosmoBounds1,CosmoBounds2a,CosmoBounds2b,CosmoBounds2c,CosmoBounds2d}
and/or new neutrino interactions \cite{Majoron1,190705425} which
open up new neutrino annihilation channels in the early universe.
In fact, the RPV framework itself
is one interesting example for a scenario that can potentially
change the cosmological picture and thus evade the cosmology bounds
on the neutrino masses. In particular, if R-Parity is spontaneously
broken \cite{CosmoBounds3,CosmoBounds4,Majoron2}, then the theory contains a massless Goldstone boson
(usually refered to as the Majoron and denoted by $J$)
and the left-handed neutrinos can decay invisibly \footnote{Note that current bounds on invisible decays of neutrinos
from solar neutrino and neutrino oscillation experiments are rather
weak \cite{190705425:27,190705425:28,190705425:29,190705425:30,190705425:31,190705425:32,190705425:33,190705425:34}.}
via $\nu_i \to \nu_j +J$ (i.e., $m_{\nu_i} > m_{\nu_j}$),
thus evading the critical density argument against
MeV-scale neutrino and, furthermore, possibly evading
the BBN constraints due to the new annihilation channel
$\nu \nu \to JJ$.
Other interesting BSM scenarios that can avoid or
significantly alleviate the cosmological
bounds on the neutrino masses can be found e.g., in the “neutrinoless 
universe”
of \cite{CosmoBounds6} and in \cite{CosmoBounds7,CosmoBounds8,CosmoBounds9} in which new physics in the neutrino 
sector was assumed.

Let us furthermore comment on the neutrinoless
double-beta ($0\nu\beta\beta$) bounds on the effective electron-neutrino
mass.
In the RPV SUSY framework, the $0\nu\beta\beta$ decay amplitude receives
additional contributions from SUSY
particles (sleptons, charginos, neutralinos, squarks and gluinos),
which do not involve the Majorana neutrino exchange mechanism, see e.g.,
\cite{nuless1,nuless2}.
Thus, any physics output of the $0\nu\beta\beta$ decay depends
on the underlying assumption and/or SUSY parameter space. In particular,
for a destructive interference between the neutrino exchange mechanism
and the
pure SUSY mechanisms, it seems plausible to
find regions in parameter space, where no constraints from
$0\nu\beta\beta$ decay
could be derived on the effective electron-neutrino mass \cite{nuless2}.

    \item[\underline{\bf Higgs signals}]
    For each point/model in our RPV SUSY parameter-space we calculate all the Higgs signal strengths in Table \ref{tab:Higgs-filters} and require them to agree with the measured ones at the 
    $2\sigma$ level.
\end{description}
    
\end{itemize}

\begin{table}
    \caption{\label{tab:Input-parameter-ranges}
    Initial input parameter ranges for the free-parameters in the numerical simulations. See also text.}
    \medskip{}
    \centering{}
    \begin{tabular}{ccc}
        \toprule 
         & Range & \tabularnewline
        \midrule
        \addlinespace
        $\delta_{\epsilon}$ & $\left[0,0.5\right]$ & \tabularnewline
        \addlinespace
        $\mu$ & $\left[90,1000\right]$ &  {[}GeV{]}\tabularnewline
        \addlinespace
        $M_{1}$ & $\left[100,2500\right]$ &  {[}GeV{]}\tabularnewline
        \addlinespace
        $M_{2}$ & $\left[100,2500\right]$ &  {[}GeV{]}\tabularnewline
        \addlinespace
        $t_{\beta}$ & $\left[2,30\right]$ & \tabularnewline
        \addlinespace
        $\delta_{B}$ & $\left[0,0.5\right]$ &  \tabularnewline
        \addlinespace
        $m_{A}$ & $\left[1000,10000\right]$ &   {[}GeV{]}\tabularnewline
        \addlinespace
        $m_{\tilde{\nu}_{\tau}}$ & $\left[200,800\right]$ &  {[}GeV{]}\tabularnewline
        \addlinespace
        $m_{\tilde{q}}$ & $\left[1000,8000\right]$ &  {[}GeV{]}\tabularnewline
        \addlinespace
        $\tilde{A}$ & $\left[0,4000\right]$ &  {[}GeV{]}\tabularnewline
        \addlinespace
        $m_{\tilde{b}_{RR}}$ & $\left[2000,5000\right]$ &  {[}GeV{]}\tabularnewline
        \addlinespace
        $m_{\tilde{\tau}_{RR}}$ &  $\left[1000,5000\right]$ &  {[}GeV{]}\tabularnewline
        \bottomrule
        \addlinespace
    \end{tabular}
\end{table}

\subsection{\label{subsec:Higgs decays to Gauginos}Higgs decays to Gauginos}

We study here the pure BRPV Higgs decays $h\rightarrow\nu_{\tau}\tilde{\chi}_{2}^{0}$ and $h\rightarrow\tau^{\pm}\chi_{2}^{\mp}$, see also \cite{RPVHiggsValle,RPVHiggsRosiek1} (for another interesting variation of RPV Higgs decays to gauginos see \cite{UnusualRPVHiggs}). Depending on the scenario under consideration, 
we require $m_{\tilde{\chi}_{2}^{0}}<125$~GeV and/or $m_{\chi_{2}^{\mp}}\lesssim125$ GeV, 
in which case the BRPV decays $h\rightarrow\nu_{\tau}\tilde{\chi}_{2}^{0}$ and/or $h\rightarrow\tau^{\pm}\chi_{2}^{\mp}$ are kinematically open, respectively 
(also adding them to the total Higgs width $\Gamma^{h}$). 

We consider four BRPV scenarios for the parameter space associated with the gaugino
sector:
\begin{description}

    \item[\texttt{S1A}]\:A gaugino-like scenario with $M_{2}\ll\mu$ \cite{NearlyDegChargNeut}, 
    and nearly degenerate 2nd lightest neutralino and chargino with a mass lighter than the Higgs mass: $m_{\tilde{\chi}_{2}^{0}} \simeq m_{\chi_{2}^{\mp}} < 125$ GeV. In this case, both 
    decays $h\rightarrow\nu_{\tau}\tilde{\chi}_{2}^{0}$ and 
    $h\rightarrow\tau^{\pm}\chi_{2}^{\mp}$ are kinematically allowed.
    \item[\texttt{S1B}]\:A higgsino-like scenario  with $\mu\ll M_{2}$ \cite{NearlyDegChargNeut},
    and nearly degenerate 2nd lightest neutralino and chargino with a mass lighter than the Higgs mass: $m_{\tilde{\chi}_{2}^{0}} \simeq m_{\chi_{2}^{\mp}} < 125$ GeV. In this case also, both 
    decays $h\rightarrow\nu_{\tau}\tilde{\chi}_{2}^{0}$ and 
    $h\rightarrow\tau^{\pm}\chi_{2}^{\mp}$ are kinematically open.
    \item[\texttt{S2}]\: No degeneracy in the gaugino sector with $m_{\tilde{\chi}_{2}^{0}}<125$~GeV 
    and $m_{\chi_{2}^{\mp}}> 125$ GeV, so that only the decay channel
    $h\rightarrow\nu_{\tau}\tilde{\chi}_{2}^{0}$ is kinematically open.
    \item[\texttt{S3}] \: No degeneracy in the gaugino sector with both 
    $m_{\tilde{\chi}_{2}^{0}},~m_{\chi_{2}^{\mp}} < 125$ GeV 
    and a significant branching fraction in the neutralino channel $h\rightarrow\nu_{\tau}\tilde{\chi}_{2}^{0}$: 
    $BR(h\rightarrow\nu_{\tau}\tilde{\chi}_{2}^{0}) \gtrsim10\%$ and a kinematically open 
    $h\rightarrow\tau^{\pm}\chi_{2}^{\mp}$ decay with much smaller rate.
\end{description}

\begin{figure}
    \begin{centering}
    \includegraphics[width=0.80\textwidth]{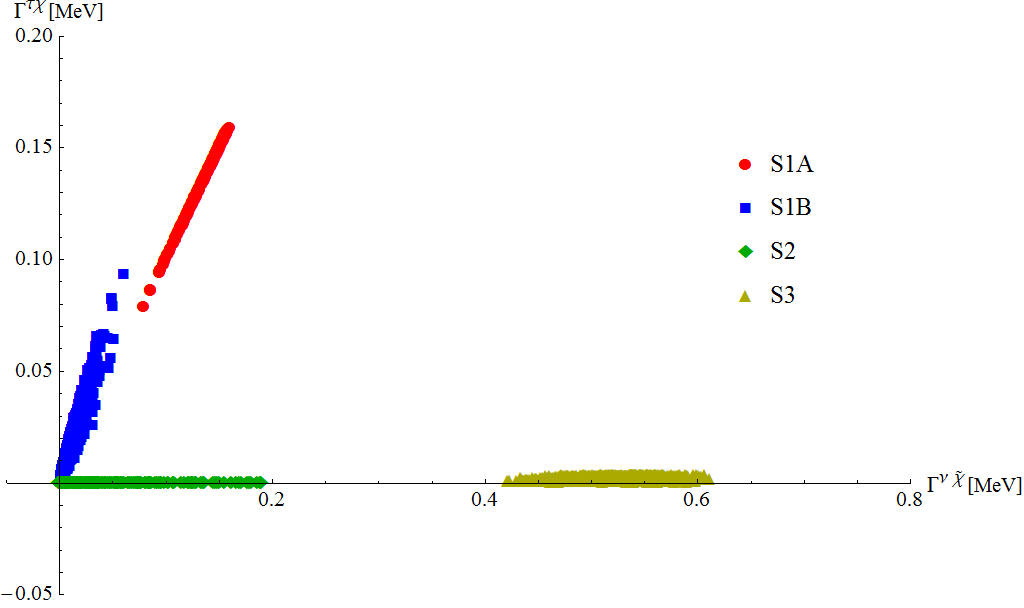}
    \par\end{centering}
    \caption{\label{fig:Scatter-plot-of-chitau-inv}
        A scatter plot
        in the $\left(\Gamma^{\nu\tilde{\chi}}\text{[MeV],\ensuremath{\Gamma^{\tau\chi}}\text{[MeV]}}\right)$
        plane for the four proposed BRPV scenarios: \texttt{S1A}, \texttt{S1B}, \texttt{S2} and \texttt{S3}, see text.}
\end{figure}

We give in Fig.~\ref{fig:Scatter-plot-of-chitau-inv} a scatter plot
of the surviving model configurations in the 
$\Gamma^{\nu \tilde{\chi}}$--$\Gamma^{\tau \chi}$ plane 
for the above four BRPV scenarios, where $\Gamma^{\nu \tilde{\chi}} = \Gamma(h\rightarrow\nu_{\tau}\tilde{\chi}_{2}^{0})$ 
and  $\Gamma^{\tau \chi}=\Gamma(h\rightarrow\tau^{\pm}\chi_{2}^{\mp})$.
We can see that within the two \texttt{S1} scenarios, \texttt{S1A} yields 
larger decay rates in both channels $h\rightarrow\nu_{\tau}\tilde{\chi}_{2}^{0}$ and 
$h\rightarrow\tau^{\pm}\chi_{2}^{\mp}$, in particular, reaching 
a width $\Gamma^{\nu \tilde{\chi}} \sim \Gamma^{\tau \chi} \sim 0.1 - 0.2$ MeV.
In the \texttt{S2} scenario we expect 
a BRPV signal only in the $h\rightarrow\nu_{\tau}\tilde{\chi}_{2}^{0}$ channel 
($h\rightarrow\tau^{\pm}\chi_{2}^{\mp}$ is kinematically closed, see 
above), which can also reach a width of $\Gamma^{\nu \tilde{\chi}} \sim 0.2$ MeV. 
Finally, we see that the \texttt{S3} scenario is expected to give the largest 
BRPV decay rate in the neutralino channel $h\rightarrow\nu_{\tau}\tilde{\chi}_{2}^{0}$, 
reaching $\Gamma(h\rightarrow\nu_{\tau}\tilde{\chi}_{2}^{0}) \sim {\cal O}(0.5)$ MeV, 
which is more than 10\% of the total SM Higgs width; in this case, the BRPV 
Higgs decay channel to a chargino, $h\rightarrow\tau^{\pm}\chi_{2}^{\mp}$ is  
effectively closed due to a limited phase-space. 
We thus see that the different \texttt{Si} SUSY scenarios that we have outlined above, probe 
different regions in the $\Gamma^{\nu \tilde{\chi}}$--$\Gamma^{\tau \chi}$ 
BRPV Higgs decays plane, 
where the cases without the $\tilde{\chi}_2^0 - \chi_2^+$ mass degeneracy 
(scenarios \texttt{S2} and \texttt{S3}) 
we obtain a better sensitivity to the neutralino channel $h\rightarrow\nu_{\tau}\tilde{\chi}_{2}^{0}$.  

In Table~\ref{tab:Input-parameters-BRPV} we list four representative benchmark models \texttt{BMi}
(i.e., sets of input parameters) which correspond to the four \texttt{Si} scenarios considered above. 
These sample benchmark models 
maximize the BRPV effect (i.e., decay rates) 
associated with the \texttt{Si} scenarios; the corresponding BRPV Higgs decay width 
into a single neutralino and a single chargino are given in 
Table~\ref{tab:BRPV-decays}.
As can be seen from Table~\ref{tab:Input-parameters-BRPV}, all four BM models require 
low $t_{\beta} \sim 2-3$. Note also that \texttt{BM3}, for which we obtain a width of 
$\Gamma(h\rightarrow\nu_{\tau}\tilde{\chi}_{2}^{0}) \sim {\cal O}(0.6)$ MeV (see Table \ref{tab:BRPV-decays}) is characterized by the hierarchy $\mu\sim M_{1}\ll M_{2}$ in 
the gaugino sector.

\begin{table}[htb]
    \caption{\label{tab:Input-parameters-BRPV}
    Input parameters for the
    selected benchmark models: \texttt{BM1A}, \texttt{BM1B}, \texttt{BM2} and \texttt{BM3}, see text.}
    \medskip{}
    \centering{}
    \begin{tabular}{cccccc}
        \toprule 
         & \texttt{BM1A} & \texttt{BM1B} & \texttt{BM2} & \texttt{BM3} &   \tabularnewline
        \midrule
        \addlinespace
        $\delta_{\epsilon}$ & $0.04$ & $0.27$ & $0.10$ & $0.22$ & \tabularnewline
        \addlinespace
        $\mu$ & $626.54$ & $92.90$ & $220.38$ & $120.05$ &  {[}GeV{]}\tabularnewline
        \addlinespace
        $M_{1}$ & $523.19$ & $2030.48$ & $104.94$ & $130.56$ &  {[}GeV{]}\tabularnewline
        \addlinespace
        $M_{2}$ & $103.83$ & $1028.05$ & $991.55$ & $999.39$ & {[}GeV{]}\tabularnewline
        \addlinespace
        $t_{\beta}$ & $2.14$ & $2.73$ & $2.81$ & $3.15$ & \tabularnewline
        \addlinespace
        $\delta_{B}$ & $0.05$ & $0.11$ & $0.17$ & $0.10$ & \tabularnewline
        \addlinespace
        $m_{A}$ & $4467.78$ & $2558.96$ & $2710.2$ & $3162.6$ & {[}GeV{]}\tabularnewline
        \addlinespace
        $m_{\tilde{\nu}_{\tau}}$ & $291.65$ & $317.38$ & $506.78$ & $358.69$ &  {[}GeV{]}\tabularnewline
        \addlinespace
        $m_{\tilde{q}}$ & $6071.69$ & $2860.5$ & $4628.27$ & $1094.07$ & {[}GeV{]}\tabularnewline
        \addlinespace
        $\tilde{A}$ & $1537.44$ & $2842.51$ & $66.19$ & $3180.91$ &  {[}GeV{]}\tabularnewline
        \addlinespace
        $m_{\tilde{b}_{RR}}$ & $4814.49$ & $4996.39$ & $4245.07$ & $4721.63$ &  {[}GeV{]}\tabularnewline
        \addlinespace
        $m_{\tilde{\tau}_{RR}}$ & $1509.25$ & $1303.96$ & $1122.68$ & $2670.45$ &  {[}GeV{]}\tabularnewline
        \bottomrule
        \addlinespace
    \end{tabular}
\end{table}
\begin{table}[htb]
    \caption{\label{tab:BRPV-decays}
    The BRPV decay width for the selected benchmark models: \texttt{BM1A}, \texttt{BM1B}, \texttt{BM2} and \texttt{BM3}, see text.}
    \medskip{}
    \centering{}
    \begin{tabular}{cccccc}
        \toprule 
         & \texttt{BM1A} & \texttt{BM1B} & \texttt{BM2} & \texttt{BM3} &   \tabularnewline
        \midrule
        \addlinespace
        $\Gamma^{\nu\tilde{\chi}}$ & $0.159$ & $0.06$ & $0.189$ & $0.61$ & {[}MeV{]} \tabularnewline
        \addlinespace
        $\Gamma^{\tau\chi}$ & $0.158$ & $0.09$ & $0$ & $0.002$ &  {[}MeV{]}\tabularnewline
        \bottomrule
        \addlinespace
    \end{tabular}
\end{table}

Another useful handle that can help distinguish between the BRPV benchmark models is the set of 
125 GeV Higgs signals considered in section \ref{sec:Higgs-signals}. In Table~\ref{tab:Higgs-observables-BRPV} we list the predicted Higgs signal strengths for 
the selected benchmark models. Indeed, we see that large deviations of ${\cal O}(25\%)$ are expected in the \texttt{BM1A} scenario in the di-photon channels $\mu_{F\gamma\gamma}^{(gg)}$ and $\mu_{V\gamma\gamma}^{(VBF)}$, due to the contribution of the light charginos in this case (see Table~\ref{tab:SUSY-spectrum-BRPV1}). In contrast, in the \texttt{BM1B} scenario, a large effect of ${\cal O}(20\%)$ is expected in $h \to \tau \tau$ channels $\mu_{F\tau\tau}^{(gg)}$ and  $\mu_{V\tau\tau}^{(VBF)}$. Furthermore, while the \texttt{BM2} setup does not exhibit significant 
deviations from the SM, the Higgs signal strengths in the $h \to \tau^+ \tau^-$ and $h \to b \bar b$ channels, $\mu_{F\tau\tau}^{(gg)}$, $\mu_{V\tau\tau}^{(VBF)}$ and $\mu_{Vbb}^{(hV)}$, exhibit non-negligible sensitivity to the benchmark model \texttt{BM3}. It is also worth noting that in all four benchmark models $\mu_{FWW}^{(gg)}=\mu_{FZZ}^{(gg)} =0.92$ (recall that in our BRPV framework 
we have $\mu_{FWW}^{(gg)}=\mu_{FZZ}^{(gg)}$), thus saturating the $2\sigma$ lower bound in these channels (see Table~\ref{tab:Higgs-filters}).     
\begin{table}
    \caption{\label{tab:Higgs-observables-BRPV}
    The Higgs observables for the
    selected benchmark models: \texttt{BM1A}, \texttt{BM1B}, \texttt{BM2} and \texttt{BM3} (see text).}
    \medskip{}
    \centering{}
    \begin{tabular}{lcccc}
        \toprule 
         & \texttt{BM1A} & \texttt{BM1B} & \texttt{BM2} & \texttt{BM3}   \tabularnewline
        \midrule
        \addlinespace
        $\mu_{F\gamma\gamma}^{(gg)}$ & $1.24$ & $1.09$ & $0.99$ & $1.01$  \tabularnewline
        \addlinespace
        $\mu_{FZZ}^{(gg)}$ & $0.92$ & $0.92$ & $0.92$ & $0.92$  \tabularnewline
        \addlinespace
        $\mu_{FWW}^{(gg)}$ & $0.92$ & $0.92$ & $0.92$ & $0.92$ \tabularnewline
        \addlinespace
        $\mu_{F\tau\tau}^{(gg)}$ & $0.91$ & $0.77$ & $0.92$ & $0.82$ \tabularnewline
        \addlinespace
        $\mu_{F\mu\mu}^{(gg)}$ & $0.92$ & $0.97$ & $0.96$ &
        $0.96$
        \tabularnewline
        \addlinespace
        $\mu_{Vbb}^{(hV)}$ & $0.92$ & $0.98$ & $0.97$ & $0.88$ \tabularnewline
        \addlinespace
        $\mu_{V\gamma\gamma}^{(VBF)}$ & $1.24$ & $1.10$ & $1.00$ & $0.93$ \tabularnewline
        \addlinespace
        $\mu_{V\tau\tau}^{(VBF)}$ & $0.92$ & $0.78$ & $0.93$ & $0.75$\tabularnewline
        \bottomrule
        \addlinespace
    \end{tabular}
\end{table}

Finally, we wish to briefly comment on the experimental signatures of the BRPV decays  $h\rightarrow\nu_{\tau}\tilde{\chi}_{2}^{0}$ and/or $h\rightarrow\tau^{\pm}\chi_{2}^{\mp}$ 
considered in this section. 
For this purpose, we compute the subsequent 
chargino and neutralino decays in our BRPV SUSY framework. 
In particular, the 
leading decays of the 2nd lightest gauginos in the BRPV scenario are the 2-body 
$\chi_{2}^{+}\rightarrow\nu W^{+}$, $\chi_{2}^{+}\rightarrow \tau^{+} Z$ and $\tilde{\chi}_{2}^{0}\rightarrow\tau^{-} W^{+}$, $\tilde{\chi}_{2}^{0}\rightarrow \nu Z$ \cite{BRPV2bodydecays,2bodyRPVdecays1,2bodyRPVdecays2,2bodyRPVdecays3}, 
since the 3-body sfermion-mediated gaugino decays (see e.g.,  \cite{DjouadigauginodecaysMSSM,3bodyBRPVdecays1}) 
are suppressed by both an extra RPV small coupling and a heavy off-shell sfermion 
propagator (in the heavy SUSY limit used in this work). In particular,   
we find that for all the above benchmark models, the gauginos decay 
almost exclusively to final states involving the $W$-boson, 
with branching ratios
$BR(\chi_{2}^{+}\rightarrow\nu W^{+}),~BR(\tilde{\chi}_{2}^{0}\rightarrow\tau^{-} W^{+}) \gtrsim90\%$. Furthermore, these gaugino 2-body BRPV decays are prompt with a lifetime corresponding to $l\sim10^{-10}\,\mathrm{m}$, i.e., they decay within the detector. 
As a result, the expected signals for both $h\rightarrow\nu_{\tau}\tilde{\chi}_{2}^{0}$ and $h\rightarrow\tau^{\pm}\chi_{2}^{\mp}$ (after the subsequent decays of the $W$) include e.g., 
a pair of opposite charged non-diagonal leptons 
$\tau^\pm e^\mp$ 
and/or $\tau^\pm \mu^\mp$ as well as a pair of 
opposite charged $\tau$-leptons with accompanying 
missing energy: 
$h\rightarrow\tau^\pm \chi_{2}^{\mp} \rightarrow \tau^{\pm}\ell^{\mp} + \missET$ and $h\rightarrow\nu_{\tau}\tilde{\chi}_{2}^{0}\rightarrow \tau^{\pm}\ell^{\mp} + \missET$, where $\ell=e,\mu,\tau$. Let us therefore define the following decay signal:
\begin{eqnarray}
\mu_{\tau \ell + \missET} \equiv \frac{\Gamma(h \to \tau^{\pm}\ell^{\mp} + \missET)}{\Gamma(h \to \tau^{\pm}\ell^{\mp} + \missET)_{SM}} ~, \label{eq:mu-dec}
\end{eqnarray}
where the dominant underlying Higgs decay in 
the SM is:\footnote{The contribution of the decay 
$h \to ZZ^\star$ to the $\tau^\pm \tau^\mp + \missET$ signal is subdominant and has a different kinematical signature.}
\begin{eqnarray}
\Gamma(h \to \tau^{\pm}\ell^{\mp} + \missET)_{SM} = \Gamma(h \to WW^\star \to \tau^{\pm}\ell^{\mp} + \missET)_{SM}~;~\ell=e,\mu,\tau ~,
\end{eqnarray}
while in our BRPV SUSY framework we have:
\begin{eqnarray}
\Gamma(h \to \tau^{\pm}\ell^{\mp} + \missET) &=& \Gamma(h \to WW^\star \to \tau^{\pm}\ell^{\mp} + \missET) + \Gamma(h\rightarrow\tau^\pm \chi_{2}^{\mp} \rightarrow \tau^{\pm}\ell^{\mp} + \missET) 
\nonumber \\ 
&+& \Gamma(h\rightarrow\nu_{\tau}\tilde{\chi}_{2}^{0}\rightarrow \tau^{\pm}\ell^{\mp} + \missET)  ~;~\ell=e,\mu,\tau ~.
\end{eqnarray}

In particular, we have 
$\Gamma(h \to WW^\star \to \tau^{\pm}\ell^{\mp} + \missET) = \left( g_{hVV}^{RPC} \right)^2 
\Gamma(h \to WW^\star \to \tau^{\pm}\ell^{\mp} + \missET)_{SM}$ (see eq.~\eqref{eq:hVVcoupling})  
so that
\begin{eqnarray}
\mu_{\tau \ell + \missET} = \left( g_{hVV}^{RPC} \right)^2 + \frac{\Gamma(h\rightarrow\tau^\pm \chi_{2}^{\mp} \rightarrow \tau^{\pm}\ell^{\mp} + \missET) + \Gamma(h\rightarrow\nu_{\tau}\tilde{\chi}_{2}^{0}\rightarrow \tau^{\pm}\ell^{\mp} + \missET)}{\Gamma(h \to WW^\star \to \tau^{\pm}\ell^{\mp} + \missET)_{SM}} ~, \label{eq:BRPVdec}
\end{eqnarray}
where the second term in eq.~\eqref{eq:BRPVdec} above is a pure BRPV effect. 

We can thus evaluate this BPRV decay signal, $\mu_{\tau \ell + \missET}$, in our four benchmark models 
\texttt{BM1A}, \texttt{BM1B}, \texttt{BM2} and \texttt{BM3}. 
In particular, in all these benchmark models we have  
$\mu_{FWW}^{(gg)} \sim \left( g_{hVV}^{RPC} \right)^2 \sim 0.92$, whereas 
$\Gamma(h\rightarrow\tau^\pm \chi_{2}^{\mp}) +\Gamma(h\rightarrow\nu_{\tau}\tilde{\chi}_{2}^{0}) \sim 0.3, 0.15, 0.2, 0.6$ in the benchmark models 
\texttt{BM1A}, \texttt{BM1B}, \texttt{BM2} and \texttt{BM3}, respectively (see Table \ref{tab:BRPV-decays}). Furthermore, as mentioned above, in all four \texttt{BMi} we have $BR(\chi_{2}^{+}\rightarrow\nu W^{+}) \gtrsim 0.9$ 
and $BR(\tilde{\chi}_{2}^{0}\rightarrow\tau^{-} W^{+}) \gtrsim 0.9$. We thus expect 
$\Gamma(h\rightarrow\tau^\pm \chi_{2}^{\mp} \rightarrow \tau^{\pm}\ell^{\mp} + \missET) + \Gamma(h\rightarrow\nu_{\tau}\tilde{\chi}_{2}^{0}\rightarrow \tau^{\pm}\ell^{\mp} + \missET) \sim 0.015 - 0.06$ MeV depending on the benchmark model, while in the SM we have $\Gamma(h \to WW^\star \to \tau^{\pm}\ell^{\mp} + \missET)_{SM} \sim 0.01~{\rm MeV}$
(recall that $BR(W \to \ell \nu_\ell) \sim 1/9$ in any single lepton decay channel of the $W$), 
so that, overall, 
we expect that the BRPV SUSY models described above will yield:
\begin{eqnarray}
\mu_{\tau \ell + \missET} \equiv \frac{\Gamma(h \to \tau^{\pm}\ell^{\mp} + \missET)}{\Gamma(h \to \tau^{\pm}\ell^{\mp} + \missET)_{SM}} \sim 2.5 - 7 ~, \label{eq:BRPVdec2}
\end{eqnarray}
which is \textbf{\textit{several times larger than}} the signal expected in the SM or in the RPC SUSY case: 
$\mu_{\tau \ell + \missET} \sim \left( g_{hVV}^{RPC} \right)^2 \sim 0.92$.

\subsection{Higgs decay to a pair of leptons: 
$h\rightarrow\mu^+\mu^-$ and $h\rightarrow\tau^+\tau^-$}

In the RPC SUSY framework the Higgs decays to a pair of $\tau$-leptons and muons are governed by the corresponding Yukawa couplings and are sensitive to the parameters in the Higgs sector, i.e., to $\tan\beta$ 
and the pseudoscalar Higgs mass $m_A$ \cite{Decoupling} 
(at tree-level). 
In particular, in the so called decoupling limit where $m_A^2 \gg m_Z^2$, 
the Higgs decays into these channels have rates very similar to the SM rates, so that the corresponding signal strengths are expected to   
be $\mu_{F\tau\tau}^{(gg)},\mu_{F\mu\mu}^{(gg)} \to 1$. Note that the Higgs decay to a pair of $\tau$-leptons is also sensitive to the Higgs signal $\mu_{V\tau\tau}^{(VBF)}$, which is also expected to be $\mu_{V\tau\tau}^{(VBF)} \to 1$ since $\mu_V^{(VBF)}\sim 1$ (see eq.~\ref{eq:prodfac}) at decoupling \cite{Decoupling}.    

On the other hand, 
when the BRPV interactions are "turned on", additional diagrams can contribute to these decays, yielding $\delta_\epsilon \cdot \delta_B$ (see e.g., diagram (b) in  
Fig.~\ref{fig:hchrydiags}) and/or $(\delta_\epsilon)^2$ BRPV effects. 
We have performed another numerical search for models that 
maximize the BRPV effects in the decays $h\rightarrow\mu^+\mu^-$ and 
$h\rightarrow\tau^+\tau^-$, 
within the ranges of input parameters used    
in Table \ref{tab:Input-parameter-ranges} and the filters described above. 
In particular, for the case of $h\rightarrow\mu^+\mu^-$ we assume that the BRPV interactions involves the 2nd generation lepton and slepton, so that in this case we assume that 
$\delta_\epsilon$ parameterizes $\mu - \chi^+$ mixing and $\delta_B$ is responsible for 
$\tilde\nu_\mu - h$ mixing. Also, we have modified the neutrino mass bound filter in the $h\rightarrow\mu^+\mu^-$ case accordingly to $m_{\nu_{\mu}}<0.19$ MeV \cite{PDG2017}. 
We note that a better sensitivity to 
the BRPV effect in the leptonic Higgs decays,  
$h\to \tau^+\tau^-,~\mu^+\mu^-$, is obtained when the Higgs decay channels to gauginos $h \to \nu_\tau \tilde\chi^0_2$ and $h \to \tau^\pm \chi^\mp_2$ are kinematically closed, i.e., when 
$m_{\tilde\chi^0_2}, m_{\chi_2^+} > m_h$. 
\begin{table}[htb]
    \caption{\label{tab:Input-parameters-mumutautau}
    Input parameter sets for the benchmark models \texttt{BM$\mu$} and \texttt{BM$\tau$}, with $l=\mu$ 
    and $l=\tau$, respectively, for the parameters $m_{\tilde{\nu}_{l}}$ and $m_{\tilde{l}_{RR}}$, 
    see also text.}
    \medskip{}
    \centering{}
    \begin{tabular}{cccc}
        \toprule 
         & \texttt{BM$\mu$} & \texttt{BM$\tau$} &  \tabularnewline
        \midrule
        \addlinespace
        $\delta_{\epsilon}$ & 0.47 & 0.49  & \tabularnewline
        \addlinespace
        $\mu$ & 642.71 & 631.61 &  {[}GeV{]}\tabularnewline
        \addlinespace
        $M_{1}$ & 1426.05 & 1651.6 &  {[}GeV{]}\tabularnewline
        \addlinespace
        $M_{2}$ & 682.82 & 687.75 &  {[}GeV{]}\tabularnewline
        \addlinespace
        $t_{\beta}$ & 6.31 & 6.76 & \tabularnewline
        \addlinespace
        $\delta_{B}$ & 0.05 & 0.05 & \tabularnewline
        \addlinespace
        $m_{A}$ & 8981.82 & 8530.08 &  {[}GeV{]}\tabularnewline
        \addlinespace
        $m_{\tilde{\nu}_{l}}$ & 543.82 & 535.47 &  {[}GeV{]}\tabularnewline
        \addlinespace
        $m_{\tilde{q}}$ & 2210.72 & 2415.51 &  {[}GeV{]}\tabularnewline
        \addlinespace
        $\tilde{A}$ & 520.38 & 247.83 &  {[}GeV{]}\tabularnewline
        \addlinespace
        $m_{\tilde{b}_{RR}}$ & 4720.75 & 4594.09 &  {[}GeV{]}\tabularnewline
        \addlinespace
        $m_{\tilde{l}_{RR}}$ & 4249.44 & 4145.23 &  {[}GeV{]}\tabularnewline
        \bottomrule
        \addlinespace
    \end{tabular}
\end{table}

\begin{table}[htb]
    \caption{\label{tab:Higgs-observables-BMmutau}
    The Higgs signal strength observables corresponding to the benchmark models \texttt{BM$\mu$} and \texttt{BM$\tau$}, see text.}
    \medskip{}
    \centering{}
    \begin{tabular}{lcc}
        \toprule 
         & \texttt{BM$\mu$} & \texttt{BM$\tau$} \tabularnewline
        \midrule
        \addlinespace
        $\mu_{F\gamma\gamma}^{(gg)}$ & $1.00$ & $1.02$   \tabularnewline
        \addlinespace
        $\mu_{FZZ}^{(gg)}$ & $0.98$ & $1.00$   \tabularnewline
        \addlinespace
        $\mu_{FWW}^{(gg)}$ & $0.98$ & $1.00$  \tabularnewline
        \addlinespace
        $\mu_{F\tau\tau}^{(gg)}$ & $0.99$ & $0.73$  \tabularnewline
        \addlinespace
        $\mu_{F\mu\mu}^{(gg)}$ & $0.75$ & $1.01$  \tabularnewline
        \addlinespace
        $\mu_{Vbb}^{(hV)}$ & $1.00$ & $1.02$  \tabularnewline
        \addlinespace
        $\mu_{V\gamma\gamma}^{(VBF)}$ & $1.01$ & $1.02$  \tabularnewline
        \addlinespace
        $\mu_{V\tau\tau}^{(VBF)}$ & $1.00$ & $0.73$ \tabularnewline
        \bottomrule
        \addlinespace
    \end{tabular}
\end{table}

In Table \ref{tab:Input-parameters-mumutautau} we list two representative benchmark models, \texttt{BM$\tau$} and \texttt{BM$\mu$}, for which we find a substantial deviation 
from $\mu_{F\tau\tau}^{(gg)}=\mu_{V\tau\tau}^{(VBF)}=1$ and $\mu_{F\mu\mu}^{(gg)} = 1$, respectively (as mentioned earlier, the RPC SUSY effect on the 125 GeV Higgs signals and in particular on the Higgs decays to a pair of leptons is negligible in the decoupling limit considered here). The resulting Higgs signal strength values corresponding to these two models are given  
in Table \ref{tab:Higgs-observables-BMmutau}.

We see that the BRPV effects in \texttt{BM$\tau$} and \texttt{BM$\mu$} 
reduce the signal strengths in the lepton channels by 
about 25\%, yielding $\mu_{F\tau\tau}^{(gg)}\sim \mu_{V\tau\tau}^{(VBF)}\sim 0.73$ and $\mu_{F\mu\mu}^{(gg)} \sim 0.75$, respectively, 
where these deviations from unity are primarily due to the BRPV lepton-chargino mixing parameter $\delta_{\epsilon}$, since $\delta_\epsilon \gg \delta_B$ in these benchmark models (see Table~\ref{tab:Input-parameters-mumutautau}).
This is still within the current $1 \sigma$
error on the measured signal strength in $\mu_{F\tau\tau}^{(gg)}$ and $\mu_{F\mu\mu}^{(gg)}$ (see Table \ref{tab:Higgs-filters}), 
but may turn out to be an interesting signal of RPV SUSY when a precision of 5-10\% will be reached on these quantities; in particular, since 
all other Higgs decay channels are left unchanged within the benchmark model \texttt{BM$\mu$}, whereas an interesting correlation $\mu_{F\tau\tau}^{(gg)}\sim \mu_{V\tau\tau}^{(VBF)}$ is obtained in \texttt{BM$\tau$}.

\section{\label{sec:TRPV}Trilinear RPV - Numerical results}

In this section we shortly explore some of the
direct implications of TRPV interactions on the 125 GeV Higgs production and decay modes.\footnote{We do not consider the corresponding soft-breaking TRPV terms, since these will contribute at higher orders and are therefore expected to yield smaller corrections to the Higgs observables.}
In particular, 
we will consider below the four TRPV 
couplings $\lambda^{\prime}_{311},~\lambda^{\prime}_{333}$ and $\lambda_{322},~\lambda_{233}$, 
which correspond to 
new TRPV $\tilde\nu_\tau \bar d d$ and $\tilde\nu_\tau \bar b b$ and $\tilde\nu_\tau \mu^+ \mu^-$ and $\tilde\nu_\mu \tau^+ \tau^-$ interactions, respectively, allowing also BRPV effects via $\delta_B \neq 0$ and assuming (throughout this section) that $\delta_{\epsilon} \ll \delta_{B}$, i.e., neglecting BRPV effects which are proportional to $\delta_\epsilon$.\footnote{We note that 
BRPV$\times$TRPV effects via $\delta_\epsilon \neq 0$ can have other interesting implications. For example, sbottom mixing can be altered by  
a $\delta_\epsilon \cdot \lambda^\prime$ RPV term 
 $\propto v \mu s_{\beta} \cdot (\delta_{\epsilon} \cdot \lambda^{\prime}_{333})$, which 
 in turn affects the contribution of sbottom exchange at 1-loop in the $ggh$ and $\gamma \gamma h$ couplings, as well as the predicted Higgs mass (see eq.~\eqref{eq:sfermionHiggsmasscorrection}).} 
Indeed, the BRPV $\delta_B$ term mixes the Higgs with the sneutrino states, these new TRPV couplings can change the decay rates of the 125 GeV Higgs-sneutrino mixed state in the channels $h \to \bar d d, \bar b b, \mu^+ \mu^-, \tau^+ \tau^-$, so that the potential overall RPV effect is proportional to the product of the BRPV and TRPV couplings (at the amplitude level), i.e., to $\delta_B \cdot \lambda^\prime$ or $\delta_B \cdot \lambda$, as we discuss next. 

In the following numerical study, we again employ all the constraints/filters outlined in the previous
section, i.e., Higgs mass, neutrino masses and Higgs signals. Here, however, 
the additional $\lambda$ and $\lambda^\prime$ TRPV couplings give rise to new 
loop-induced contributions to the neutrino masses, 
so that the corresponding neutrino mass filters are modified accordingly. 
In particular, the leading contribution of the TRPV interactions 
to the neutrino mass arise at 1-loop and can be estimated via
(see \cite{B3_neutrinomass_constraints} for details):\footnote{There is an additional 
 1-loop BRPV$\times$TRPV contribution to the neutrino mass which is $\propto \delta_\epsilon \cdot \lambda^\prime$ and which we do not consider here, assuming that it is much smaller by virtue of $\delta_{\epsilon} \to 0$.} 
\begin{eqnarray}
\left(m_{\nu_{\tau}}\right)_{loop}^{\lambda^{\prime}_{3ii}\lambda^{\prime}_{3ii}} &\sim& \frac{3}{8\pi^2}(\lambda^{\prime}_{3ii})^2\frac{m_{q_i}^2}{{\bar m}_{{\tilde q}_i}} ~, \label{eq:TRPVloop1}
\end{eqnarray}
where $m_{q_1}=m_d,m_{q_3}=m_b$ are the $d$ and $b$-quark masses, respectively, and 
$\bar m_{\tilde q_1},\bar m_{\tilde q_3}=\bar m_{\tilde d},\bar m_{\tilde b}$ are the average masses of the sdown and the sbottom, respectively. 
Similarly, the 1-loop contributions for the $\lambda_{233}$ and $\lambda_{322}$ couplings are:
\begin{eqnarray}
\left(m_{\nu_{k}}\right)_{loop}^{\lambda_{kii}\lambda_{kii}} &\sim& \frac{1}{8\pi^2}(\lambda_{kii})^2\frac{m_{\ell_i}^2}{\bar m_{{\tilde \ell}_i}} ~,
\end{eqnarray}
where here $m_{\nu_2}=m_{\nu_\mu}$ and $m_{\nu_3}=m_{\nu_\tau}$ 
and 
$\bar m_{{\tilde \ell}_2}=\bar m_{{\tilde \mu}}$, $\bar m_{{\tilde \ell}_3}=\bar m_{{\tilde \tau}}$ 
are the corresponding average masses of the muon and $\tau$-neutrino charged slepton masses.   

We note, however, that the above 1-loop pure TRPV corrections to the neutrino masses are 
sub-dominant in the scenarios 
considered below, i.e., with a multi-TeV squarks and charged sleptons spectrum; 
the largest effect arises from the $\lambda^\prime_{333}$ coupling, since it is proportional to the b-quark mass, see eq.~\eqref{eq:TRPVloop1}.

\subsection{ The Higgs signals and $\delta_{B} \cdot \lambda^{\prime}$ RPV effects}

\begin{figure}[htb]
    \begin{centering}
    \includegraphics[width=0.8\textwidth]{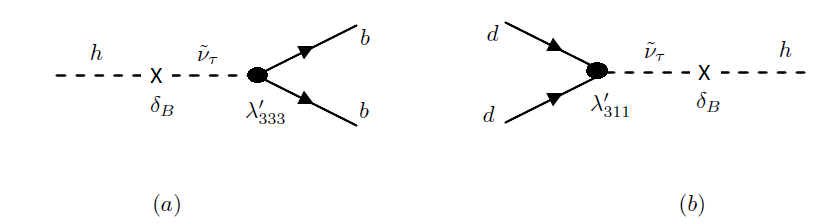}
    \par\end{centering}
    \caption{\label{fig:TRPV2}
        Tree-level diagrams (couplings) that correspond to the main $\delta_{B} \cdot \lambda^{\prime}$ effects in the Higgs decay $h\rightarrow b\overline{b}$
        (diagram (a)) and in Higgs production via $d\overline{d}$ fusion (diagram (b)). The BRPV $\delta_{B}$ insertion is denoted by X whereas the new $\lambda^{\prime}$ TRPV 
        interactions appear in bold vertices.
    }
\end{figure}

As schematically depicted in Fig.~\ref{fig:TRPV2}, when $\lambda^{\prime}_{333} \neq 0$ the Higgs coupling to bottom quarks (see also eq.~\eqref{eq:hqqcoupling}) receives a new TRPV term
proportional to $\lambda^{\prime}_{333}Z_{h3}$ (recall that 
$Z_{h3}=Z_{h3}(\delta_B)$):
\begin{align}
\Lambda_{hb\overline{b}} & =g_{b}^{SM}\left(g_{hb\overline{b}}^{RPC}+\frac{\lambda^{\prime}_{333}Z_{h3}}{\sqrt{2}g_{b}^{SM}}\right)\label{eq:hbbcouplingTRPV}\end{align}
where we have normalized the new TRPV contribution to the SM $hbb$ coupling, $g_{b}^{SM}=\frac{m_{b}}{v}$, 
and denoted the RPC $hbb$ coupling by $g_{hb\overline{b}}^{RPC} \equiv \frac{Z_{h1}}{c_{\beta}}$ 
(see also eq.~\eqref{eq:hqqcoupling}).

The new TRPV term in eq.~\eqref{eq:hbbcouplingTRPV} thus modifies (at tree-level) the Higgs 
decay $h\to b\overline{b}$:
\begin{align}
\Gamma\left(h\to b\overline{b}\right) & =3\frac{G_{F}m_{b}^{2}}{4\sqrt{2}\pi}\left(g_{hb\overline{b}}^{RPC}+\frac{\lambda^{\prime}_{333}Z_{h3}}{\sqrt{2}g_{b}^{SM}}\right)^{2}m_{h}\left(1-\frac{4m_{b}^{2}}{m_{h}^{2}}\right)^{\frac{3}{2}}\label{eq:hbbdecayTRPV} ~, 
\end{align}
and also the Higgs decays 
$h \to \gamma\gamma, gg$ at 1-loop. In particular, it modifies
the 1-loop Higgs production in the gluon-fusion channel.\footnote{The Higgs production via  
$b$-quark fusion, $b \bar b \to h$, is also modified by the extra TRPV term in eq.\eqref{eq:hbbcouplingTRPV}, but this channel is sub-dominant due to the small PDF content of 
the $b,\bar b$ quarks in the proton and is, therefore, neglected here.}

Similar to the $\lambda^\prime_{333}$ effect in the 
$h b \bar b$ coupling, when $\lambda^\prime_{311} \neq 0$ the couplings of the Higgs to a pair of $d$-quarks 
is also shifted by the term $(1/\sqrt{2})\lambda^{\prime}_{311}Z_{h3}$. In this case however, the TRPV effect is manifest by an enhanced Higgs production mechanism via $d\overline{d}$-fusion (see diagram (b) in Fig.~\ref{fig:TRPV2}); 
the corresponding TRPV effect in the Higgs decay $h\to d\overline{d}$ is not of our interest here since it is not yet measurable. Thus,
in the presence of a non-zero TRPV $\lambda^\prime_{311}$ coupling we have (for the definition of the production factors $\mu^{(P)}_i$ see eqs.~\eqref{eq:Dfac} and \eqref{eq:ggfuse}):\footnote{We set $\sigma(d \bar d \to h)_{SM}=0$ in eq.~\eqref{eq:sigmaratioTRPV}.}:
\begin{eqnarray}
\mu^{(gg+dd)}_F \equiv \frac{\sigma(gg \to h) + \sigma(d \bar d \to h)}{\sigma(gg \to h)_{SM}} = \mu_F^{(gg)} + \frac{\sigma(d \bar d \to h)}{\sigma(gg \to h)_{SM}} ~, \label{eq:sigmaratioTRPV} 
\end{eqnarray}
where the first term in eq.~\eqref{eq:sigmaratioTRPV}, $\mu_F^{(gg)}$, is the scaled gluon-fusion 
production factor in the RPV framework as defined in eq.~\eqref{eq:ggfuse} and  
calculated using eqs.~\eqref{eq:gghproductionSM}--\eqref{eq:gghproduction}.\footnote{The TRPV effect of the $d$-quark loop (via the $\lambda_{311}^\prime$ coupling) in the gluon-fusion channel is negligible, see e.g., \cite{universal_paper}.} The second term 
in eq.~\eqref{eq:sigmaratioTRPV} requires special care, in particular, since the PDF's do not cancel out when taking the ratio. It is convenient to normalize the Higgs coupling to down quarks by the SM bottom-quark yukawa, $y_{b}=\sqrt{2}g_{b}^{SM} = \sqrt{2} m_b/v$ and adopt the coupling modifier formalism (``Kappa framework"), defining (see e.g., \cite{universal_paper}): 
\begin{align}
\kappa^{TRPV}_{d} & \equiv\frac{\lambda^{\prime}_{311}Z_{h3}}{y_{b}}\label{eq:kappaTRPV} ~.
\end{align}
in which case the second term in eq.~\eqref{eq:sigmaratioTRPV} can be written as:
\begin{eqnarray}
\frac {\sigma(d \bar d \to h)}{\sigma(gg \to h)_{SM}}
 \simeq \frac{\left(\kappa^{TRPV}_{d}\right)^2 \cdot \sigma(d \bar d \to h)_{\kappa^{TRPV}_{d} =1} \cdot K_d }{\sigma(gg \to h)_{SM}^{N3LO}} \simeq 0.73 \left(\kappa^{TRPV}_{d}\right)^2  \label{eq:TRPV2ndterm} ~,
\end{eqnarray}
where $\sigma(d \bar d \to h)_{\kappa^{TRPV}_{d} =1} \simeq23.8\,\text{[pb]}$ \cite{universal_paper}, 
$\sigma(gg \to h)_{SM}^{N3LO} \simeq 48.6\,\text{[pb]}$ is 
the N3LO QCD prediction  for the gluon-fusion Higgs  production channel at the 13 TeV LHC \cite{N3LOQCD} and $K_d\simeq1.5$ is the estimated K-factor for 
the sub-process $d \bar d \to h$ with $\kappa^{TRPV}_{d} =1$ \cite{Kfactor}.

In Tables~\ref{tab:Input-parameters-TRPV} and ~\ref{tab:Higgs-observables-TRPV} we list the input parameters and the resulting Higgs signal strength observables for two benchmark models \texttt{BM$\lambda^{\prime}_{333}$} and \texttt{BM$\lambda^{\prime}_{311}$}, setting $\lambda^{\prime}_{333}\sim 0.5$ or  $\lambda^{\prime}_{311}\sim1$, respectively,  which correspond to the (conservative)\footnote{The bounds on the TRPV parameters $\lambda,\lambda^{\prime}$ scale as $1/m_{\tilde f_R}$ \cite{TRPVbounds} and can, therefore, be relaxed for $m_{\tilde f_R} > 1$ TeV (as assumed here). These bounds are also model-dependent, see e.g., \cite{TRPVboundsdiscussion}.} upper bounds for squark masses above 1 TeV, see \cite{TRPVbounds}.
The benchmark model \texttt{BM$\lambda^{\prime}_{333}$} 
has been chosen to maximize the TRPV effect in the Higgs decay $h\to b\overline{b}$, while in 
\texttt{BM$\lambda^{\prime}_{311}$} the ratio in eq.~\eqref{eq:TRPV2ndterm} and, 
therefore, the Higgs production channel via $d \bar d$-fusion are maximized.

Summarizing our results in 
Tables~\ref{tab:Input-parameters-TRPV} and~\ref{tab:Higgs-observables-TRPV}, we note that: 
\begin{itemize}
    \item The \texttt{BM$\lambda^{\prime}_{333}$} scenario exhibits only a mild enhancement in the $h\to b\overline{b}$ channel:  $\mu_{Vbb}^{(hV)}=1.04$. This implies that the Higgs decay channel 
    $h \to b \bar b$ is dominated by the $b$-quark yukawa coupling, $y_{b}$, so that 
    the new TRPV term in eq.~\eqref{eq:hbbcouplingTRPV} can be neglected in this case.
    On the other hand, the diphoton channels in the \texttt{BM$\lambda^{\prime}_{333}$} model are significantly enhanced: $\mu_{F\gamma\gamma}^{(gg)}\sim\mu_{V\gamma\gamma}^{(VBF)}\sim1.25$, primarily due to the light chargino spectrum in this case (see Table~\ref{tab:SUSY-spectrum-TRPV1}). Also, the vector boson decay channels saturate 
    their $2 \sigma$ lower bound in the \texttt{BM$\lambda^{\prime}_{333}$} scenario, i.e., $\mu_{FZZ}^{(gg)}\sim\mu_{FWW}^{(gg)}\sim0.92$. 
    \item In the \texttt{BM$\lambda^{\prime}_{311}$} scenario we have 
    $\kappa^{TRPV}_{d} \sim 1.34$, so that 
    the enhancement in the $d \bar d \to h$ production channel, see eq.~\eqref{eq:TRPV2ndterm}, causes  
    the (previously) gluon-fusion Higgs production mode 
    to be roughly doubled, i.e, we find $\mu^{(gg+dd)}_F \simeq 2.3$ in eq.~\eqref{eq:sigmaratioTRPV}.  
    On the other hand, the total Higgs decay width becomes larger due to the new enhanced $h \to d \bar d$ channel, so that the individual Higgs branching ratios are decreased. The net effect is an enhancement in what was previously the gluon-fusion initiated channels and a decrease in the  
    vector-boson initiated signals ($\mu_{Vjj}$). For example, a ${\cal O}(50\%)$ enhancement is found in $pp \to h \to \tau^+ \tau^-$ ($\mu_{F\tau\tau}^{(gg)}\sim1.5$, see Table ~\ref{tab:Higgs-observables-TRPV}), partly due to the large $t_{\beta} \sim 16$ in this model and 
    a ${\cal O}(50\%)$ suppression is predicted in this case in the VBF di-photon channel, i.e., 
    $\mu_{V\gamma \gamma}^{(VBF)} \sim 0.5$.  
\end{itemize}

We note that the TRPV Higgs coupling to the $d$-quarks, $\lambda^\prime_{311}$, also contributes to the $hV$ production channel via a t-channel $d-$quark exchange diagram, 
$d \bar d \to hV$, and therefore modifies the Higgs production factor in this channel:
\begin{eqnarray}
\mu_V^{(hV+dd)} \equiv \frac{\sigma(q \bar q \to V \to h V) + \sigma(d \bar d \to h V)}{\sigma(q \bar q \to V \to h V)_{SM}} = \mu_V^{(hV)} + \frac{\sigma(d \bar d \to hV)}{\sigma(q \bar q \to V \to h V)_{SM}} ~, \label{eq:sigmahVTRPV} 
\end{eqnarray}
where $\mu_V^{(hV)} = \left( g_{hVV}^{RPC} \right)^2$ is the 
$hV$ production factor in the RPC limit and also in our 
BRPV scenario (since the $hVV$ 
SUSY coupling is not changed in the BRPV case within the no-VEV basis $\left< v_{\tilde\nu} \right>$, see eq.~\eqref{eq:prodfac}). 
Following the above prescription, here also 
we can define the scaled t-channel $hV$ cross-section via:
\begin{eqnarray}
\sigma(d \bar d \to hV) = \left(\kappa^{TRPV}_{d}\right)^2 \cdot \sigma(d \bar d \to h V)_{\kappa^{TRPV}_{d} =1} ~,
\end{eqnarray}
where, using {\sc MadGraph5\_aMC@NLO} \cite{MG5}, we find 
(see also \cite{universal_paper}):
\begin{eqnarray}
\frac{\sigma(d \bar d \to hV)_{\kappa^{TRPV}_{d} =1}}{\sigma(q \bar q \to V \to h V)_{SM}} \sim 0.05 ~.
\end{eqnarray}

Thus, the overall change expected in the $hV$ production channel signal 
due to $\lambda^\prime_{311} \neq 0$ is:
\begin{eqnarray}
\mu_V^{(hV+dd)} \simeq\left(g_{hVV}^{RPC}\right)^{2}+0.05 \cdot (\kappa^{TRPV}_{d})^2 ~, \label{eq:TRPVhVproduction}
\end{eqnarray}
which enters only in the $pp \to hV \to V b \bar b$ channel, i.e., $\mu_{Vbb}^{(hV)} \to \mu_{Vbb}^{(hV+dd)}$,
and was taken into account in the above analysis, i.e., in Table~\ref{tab:Higgs-observables-TRPV}.  

Finally, it is also interesting to note that the effect of a new TRPV 
$hdd$ coupling may also show up in the Higgs pair-production 
channel $pp \to hh$, as was suggested in a different context
in \cite{universal_paper}. 

\begin{table}
    \caption{\label{tab:Input-parameters-TRPV}
    Input parameters for the
    selected benchmark models \texttt{BM$\lambda^{\prime}_{333}$} and \texttt{BM$\lambda^{\prime}_{311}$}. See also text.}
    \medskip{}
    \centering{}
    \begin{tabular}{cccc}
        \toprule 
         & \texttt{BM$\lambda^{\prime}_{333}$} & \texttt{BM$\lambda^{\prime}_{311}$} &  \tabularnewline
        \midrule
        \addlinespace
        $\delta_{\epsilon}$ & $0$ & $0$ &  \tabularnewline
        \addlinespace
        $\mu$ & $202.46$ & $556.34$ &   {[}GeV{]}\tabularnewline
        \addlinespace
        $M_{1}$ & $759.74$ & $1747.98$ &   {[}GeV{]}\tabularnewline
        \addlinespace
        $M_{2}$ & $251.55$ & $1589.49$ & {[}GeV{]}\tabularnewline
        \addlinespace
        $t_{\beta}$ & $2.77$ & $16.59$ & \tabularnewline
        \addlinespace
        $\delta_{B}$ & $0.11$ & $0.45$ & \tabularnewline
        \addlinespace
        $m_{A}$ & $2150.46$ & $1508.96$ & {[}GeV{]}\tabularnewline
        \addlinespace
        $m_{\tilde{\nu}_{\tau}}$ & $768$ & $723.75$ &  {[}GeV{]}\tabularnewline
        \addlinespace
        $m_{\tilde{q}}$ & $3461.04$ & $2008.27$ &  {[}GeV{]}\tabularnewline
        \addlinespace
        $\tilde{A}$ & $953.94$ & $2.89$ & {[}GeV{]}\tabularnewline
        \addlinespace
        $m_{\tilde{b}_{RR}}$ & $2764.42$ & $2421.53$ &  {[}GeV{]}\tabularnewline
        \addlinespace
        $m_{\tilde{\tau}_{RR}}$ & $2357.42$ & $3693.50$ &  {[}GeV{]}\tabularnewline
        \bottomrule
        \addlinespace
    \end{tabular}
\end{table}

\begin{table}
    \caption{\label{tab:Higgs-observables-TRPV}
    The Higgs observables for the
    selected benchmark models \texttt{BM$\lambda^{\prime}_{333}$} and \texttt{BM$\lambda^{\prime}_{311}$}. See also text.}
    \medskip{}
    \centering{}
    \begin{tabular}{lcc}
        \toprule 
         & \texttt{BM$\lambda^{\prime}_{333}$} & \texttt{BM$\lambda^{\prime}_{311}$}   \tabularnewline
        \midrule
        \addlinespace
        $\mu_{F\gamma\gamma}^{(gg)}$ & $1.26$ & $1.11$  \tabularnewline
        \addlinespace
        $\mu_{FZZ}^{(gg)}$ & $0.92$ & $1.09$   \tabularnewline
        \addlinespace
        $\mu_{FWW}^{(gg)}$ & $0.92$ & $1.09$  \tabularnewline
        \addlinespace
        $\mu_{F\tau\tau}^{(gg)}$ & $0.93$ & $1.51$  \tabularnewline
        \addlinespace
        $\mu_{F\mu\mu}^{(gg)}$ & $0.93$ & $1.51$
        \tabularnewline
        \addlinespace
        $\mu_{Vbb}^{(hV)}$ & $1.04$ & $0.71$ \tabularnewline
        \addlinespace
        $\mu_{V\gamma\gamma}^{(VBF)}$ & $1.27$ & $0.48$ \tabularnewline
        \addlinespace
        $\mu_{V\tau\tau}^{(VBF)}$ & $0.94$ & $0.65$ \tabularnewline
        \bottomrule
        \addlinespace
    \end{tabular}
\end{table}

\subsection{ The Higgs signals and $\delta_{B} \cdot \lambda$ RPV effects}

When $\lambda_{322} \neq 0$ or $\lambda_{233} \neq 0$, the Higgs decay channels $h \to \mu^+ \mu^-$ 
or $h \to \tau^+ \tau^-$  are altered, respectively (here also we consider one TRPV coupling at a time). These effects are similar to that depicted in 
diagram Fig.~\ref{fig:TRPV2}(a), replacing $\lambda^{\prime} \to \lambda$ and the outgoing b-quarks with the corresponding leptons. 
Recall that the TRPV parameters $\lambda_{ijk}$ are anti-symmetric in their first two indices. We thus 
restrict ourselves to a one parameter scheme considering one sneutrino type at a time: 
for the $\lambda_{322} \neq 0 $ case we assume 
BRPV via $\tilde\nu_\tau - h$ mixing, whereas when $\lambda_{233} \neq 0 $ the BRPV is mediated via  $\tilde\nu_\mu - h$ mixing. Accordingly, 
in the $\tilde\nu_\tau - h$ mixing BRPV scenario 
we apply the neutrino mass bound $m_{\nu_{\tau}}<18.2$ MeV on the tau-neutrino and in the $\tilde\nu_\mu - h$ mixing case 
we apply the bound $m_{\nu_{\mu}}<0.19$ MeV on the muon-neutrino. 
We do not consider here the possible implications of the $\lambda$ TRPV couplings on the flavor violating Higgs decay $h\to \tau\mu$, which was studied in detail in \cite{Arhrib}. 

As in the $\lambda^\prime$ TRPV case, in the presence of $\lambda_{233} \neq 0$ or $\lambda_{322} \neq 0$, 
the coupling of the Higgs-sneutrino mixed state to $\tau$'s or muons
receives a new TRPV term
$\propto \lambda_{233} Z_{h3}$ or $\propto \lambda_{322} Z_{h3}$, respectively (the RPC couplings $g_{h l l}^{RPC}$ are defined in eq.~\eqref{eq:hllcoupling}):
\begin{align}
\Lambda_{h\tau\tau} & =g_{\tau}^{SM}\left(g_{h\tau\tau}^{RPC}+\frac{\lambda_{233}Z_{h3}}{\sqrt{2}g_{\tau}^{SM}}\right)\label{eq:htautaucouplingTRPV}  ~, \end{align}
\begin{align}
\Lambda_{h\mu\mu} & =g_{\mu}^{SM}\left(g_{h\mu\mu}^{RPC}+\frac{\lambda_{322}Z_{h3}}{\sqrt{2}g_{\mu}^{SM}}\right)\label{eq:hmumucouplingTRPV} ~, \end{align}
which directly modifies (at tree-level) the Higgs decays $h \to \tau^+ \tau^-$ or $h \to \mu^+ \mu^-$ and also mildly modifies 
the 1-loop $\tau$ or $\mu$ exchanges in $h \to \gamma \gamma$. 

In Tables~\ref{tab:Input-parameters-TRPVl} and~\ref{tab:Higgs-observables-TRPVl} we list the input parameters and the resulting Higgs signal strength observables for two benchmark models \texttt{BM$\lambda_{233}$} and \texttt{BM$\lambda_{322}$}, setting $\lambda_{233}=0.7$ or $\lambda_{322}=0.7$ in the superpotential, which are the (conservative) upper 
bounds for $m_{\tilde\tau_R} > 1$ TeV and $m_{\tilde\mu_R} > 1$ TeV, respectivly, see \cite{TRPVbounds}.
The \texttt{BM$\lambda_{233}$} model 
has been chosen 
to maximize the TRPV effect in the Higgs decay 
$h \to \tau^+ \tau^-$, while 
\texttt{BM$\lambda_{322}$}
maximizes the TRPV effect in $h \to \mu^+ \mu^-$; 
both within the $2\sigma$ upper bounds on the corresponding Higgs signals, see Table \ref{tab:Higgs-filters}.
In the \texttt{BM$\lambda_{322}$} scenario we have also checked that with $\lambda_{322}=0.7$ the contribution to the muon anomalous magnetic moment, $(g-2)_{\mu}$, lies within the experimental bound \cite{gminus2muonPDG}, see also \cite{gminus2muonTRPV}.

Summarizing our findings and the results in Tables~\ref{tab:Input-parameters-TRPVl} and~\ref{tab:Higgs-observables-TRPVl} we find that: 
\begin{itemize}
     \item A better sensitivity to 
these TRPV couplings via 
$h\to \tau^+\tau^-,~\mu^+\mu^-$ is obtained when the Higgs decay channels to gauginos are kinematically closed, i.e., when the 2nd lightest gauginos are heavier than the lightest Higgs, as is the case in both \texttt{BM$\lambda_{233}$} and \texttt{BM$\lambda_{322}$} models, see Table \ref{tab:SUSY-spectrum-TRPV2}. 
    \item As expected, in the \texttt{BM$\lambda_{233}$} case the Higgs signals involving $\tau$ decays are significantly 
    enhanced by the new TRPV coupling: $\mu_{F\tau\tau}^{(gg)}\sim\mu_{V\tau\tau}^{(VBF)}\sim1.85$ (we have explicitly checked that the corresponding signal strengths in the RPC SUSY limit are close to unity, $\mu_{F\tau\tau,V \tau\tau}(\lambda_{233}=0) \sim 1$ due to decoupling). 
    This is in contrast to the BRPV scenario \texttt{BM$\tau$} with $\delta_\epsilon \sim 0.5$ discussed in the previous section, where the signal strength factors in the $\tau \tau$-channels were suppressed (see Tables ~\ref{tab:Input-parameters-mumutautau}-\ref{tab:Higgs-observables-BMmutau}). The rest of the Higgs signals in the \texttt{BM$\lambda_{233}$} scenario are suppressed with respect to the SM and to the RPC SUSY case. In particular, in the vector-boson Higgs decay channels they saturate their lower $2\sigma$ bound, i.e., $\mu_{FZZ}^{(gg)}\sim\mu_{FWW}^{(gg)}\sim0.92$ and in the $pp \to h \to \mu^+ \mu^-$ channel we have $\mu_{F\mu\mu}^{(gg)}\sim 0.94$, 
    primarily due to the enlarged total Higgs decay width 
    thereby decreasing the $BR(h \to \mu^+\mu^-)$.  
    \item The \texttt{BM$\lambda_{322}$} scenario exhibits a large enhancement in the Higgs decay to muons, saturating the upper bound: $\mu_{F\mu\mu}^{(gg)}\sim1.96$, while keeping the rest of the Higgs signals around unity, which is the value expected in the decoupling RPC SUSY limit 
    and in the SM (we again verified that $\mu_{F\mu\mu}^{(gg)}(\lambda_{322}=0)\sim 1 $, as expected due to decoupling in the RPC SUSY spectrum in this case).
    Here also the enhanced signal strength in the 
    $h \to \mu^+ \mu^-$ channel is in contrast to the BRPV scenario \texttt{BM$\mu$} with 
    $\delta_\epsilon \sim 0.5$, for which we found $\mu_{F\mu\mu}^{(gg)}\sim0.75$ (see Tables ~\ref{tab:Input-parameters-mumutautau}-\ref{tab:Higgs-observables-BMmutau}).       
\end{itemize}

\begin{table}
    \caption{\label{tab:Input-parameters-TRPVl}
    Input parameters for the
    selected benchmark models \texttt{BM$\lambda_{233}$} and  \texttt{BM$\lambda_{322}$}, with $l=\mu$ and $l=\tau$, respectively, 
    for the parameters $m_{\tilde{\nu}_{l}}$ and $m_{\tilde{l}_{RR}}$.}
    \medskip{}
    \centering{}
    \begin{tabular}{cccc}
        \toprule 
         &  \texttt{BM$\lambda_{233}$} & \texttt{BM$\lambda_{322}$} &   \tabularnewline
        \midrule
        \addlinespace
        $\delta_{\epsilon}$ &  $0$ & $0$ & \tabularnewline
        \addlinespace
        $\mu$ &  $958.82$ & $270.48$ &  {[}GeV{]}\tabularnewline
        \addlinespace
        $M_{1}$ &  $593.21$ & $290.19$ &  {[}GeV{]}\tabularnewline
        \addlinespace
        $M_{2}$ &  $1355.12$ & $1222.63$ & {[}GeV{]}\tabularnewline
        \addlinespace
        $t_{\beta}$ &  $4.35$ & $2.72$ & \tabularnewline
        \addlinespace
        $\delta_{B}$ &  $0.03$ & $0.02$ & \tabularnewline
        \addlinespace
        $m_{A}$ &  $2141.48$ & $5007.63$ & {[}GeV{]}\tabularnewline
        \addlinespace
        $m_{\tilde{\nu}_{l}}$ &  $218.16$ & $718.52$ &  {[}GeV{]}\tabularnewline
        \addlinespace
        $m_{\tilde{q}}$ &  $2591.04$ & $2782.38$ & {[}GeV{]}\tabularnewline
        \addlinespace
        $\tilde{A}$ &  $95.18$ & $1772.84$ &  {[}GeV{]}\tabularnewline
        \addlinespace
        $m_{\tilde{b}_{RR}}$ &  $4703.45$ & $2381.95$ &  {[}GeV{]}\tabularnewline
        \addlinespace
        $m_{\tilde{l}_{RR}}$ &  $3133.34$ & $2371.34$ &  {[}GeV{]}\tabularnewline
        \bottomrule
        \addlinespace
    \end{tabular}
\end{table}

\begin{table}
    \caption{\label{tab:Higgs-observables-TRPVl}
    The Higgs observables for the
    selected benchmark models  \texttt{BM$\lambda_{233}$} and  \texttt{BM$\lambda_{322}$}. See also text.}
    \medskip{}
    \centering{}
    \begin{tabular}{lcc}
        \toprule 
         &  \texttt{BM$\lambda_{233}$} & \texttt{BM$\lambda_{322}$}   \tabularnewline
        \midrule
        \addlinespace
        $\mu_{F\gamma\gamma}^{(gg)}$ &  $0.94$ & $1.04$  \tabularnewline
        \addlinespace
        $\mu_{FZZ}^{(gg)}$ & $0.92$ & $0.99$  \tabularnewline
        \addlinespace
        $\mu_{FWW}^{(gg)}$ &  $0.92$ & $0.99$ \tabularnewline
        \addlinespace
        $\mu_{F\tau\tau}^{(gg)}$ &  $1.85$ & $0.99$ \tabularnewline
        \addlinespace
        $\mu_{F\mu\mu}^{(gg)}$ & $0.94$ & $1.96$
        \tabularnewline
        \addlinespace
        $\mu_{Vbb}^{(hV)}$ &  $0.94$ & $1.00$ \tabularnewline
        \addlinespace
        $\mu_{V\gamma\gamma}^{(VBF)}$ &  $0.95$ & $1.05$ \tabularnewline
        \addlinespace
        $\mu_{V\tau\tau}^{(VBF)}$ &  $1.86$ & $1.00$\tabularnewline
        \bottomrule
        \addlinespace
    \end{tabular}
\end{table}

\subsection{TRPV case - final note}

The TRPV benchmark models considered in section \ref{sec:TRPV} are by no means unique, in the sense that observable TRPV effects in the Higgs signals considered above are possible within a wide range of the SUSY parameter space and, in particular, of the TRPV and BRPV couplings, 
e.g., with significantly smaller values of the TRPV parameters. 
To demonstrate that, we consider below the Higgs signal in the $pp \to h \to \mu^+ \mu^-$ channel, $\mu_{F \mu\mu}^{(gg)}$, 
within the $\lambda_{322} \neq 0$ TRPV scenario, 
this time treating $\lambda_{322}$ as a free parameter in 
the range $\lambda_{322} \in [0,0.7]$ and 
fixing $m_A=2$ TeV with either $t_\beta=2$ or $t_\beta=30$. The 
rest of the input parameters (apart from the BRPV parameter $\delta_\epsilon$ which is again set to zero in accordance with the working assumption of section \ref{sec:TRPV}) are varied in the "standard" ranges given in Table~\ref{tab:Input-parameter-ranges}, i.e., here also the BRPV Higgs-sneutrino mixing parameter, $\delta_B$, is varied in the range $[0,0.5]$. We again apply 
all the filters that were used in the previous sections including the $95\%$ CL bound on this channel, i.e., $\mu_{F \mu \mu}^{(gg)} \leq 1.96$ in Table~\ref{tab:Higgs-filters}.

We define the RPV effect as the ``distance" from the RPC 
expectation:
\begin{align}
\Delta \mu_{F\mu\mu} & \equiv\frac{|\mu_{F\mu\mu}^{TRPV}-\mu_{F\mu\mu}^{RPC}|}{\mu_{F\mu\mu}^{RPC}}\label{eq:RPCdistance}
\end{align}
where $\mu_{F\mu\mu}^{TRPV} \equiv \mu_{F\mu\mu}^{(gg)}(\lambda_{322},\delta_B)$ and $\mu_{F\mu\mu}^{RPC}=\mu_{F\mu\mu}^{(gg)}(\lambda_{322}=0,\delta_B=0)$. We recall again that, since we work in the decoupling SUSY limit, we have $\mu_{F\mu\mu}^{RPC} \simeq 
\mu_{F\mu\mu}^{SM} \simeq 1$.    

\begin{figure}[htb]
\begin{subfigure}{.5\textwidth}
  \centering
  \includegraphics[width=.8\linewidth]{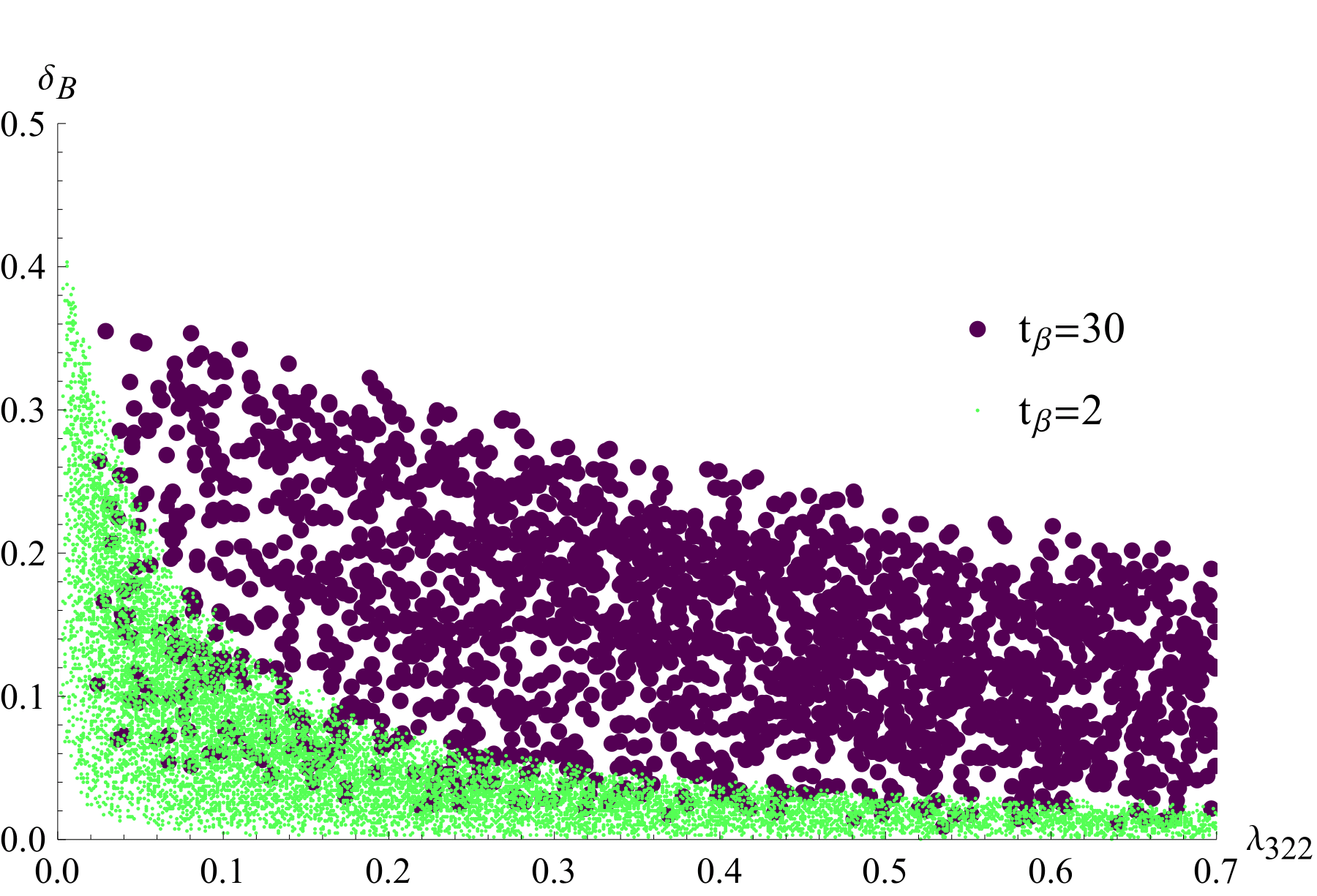}
  \caption{$\Delta\mu_{F\mu\mu}>0.5$}
  \label{fig:RPC0p5}
\end{subfigure}%
\begin{subfigure}{.5\textwidth}
  \centering
  \includegraphics[width=.8\linewidth]{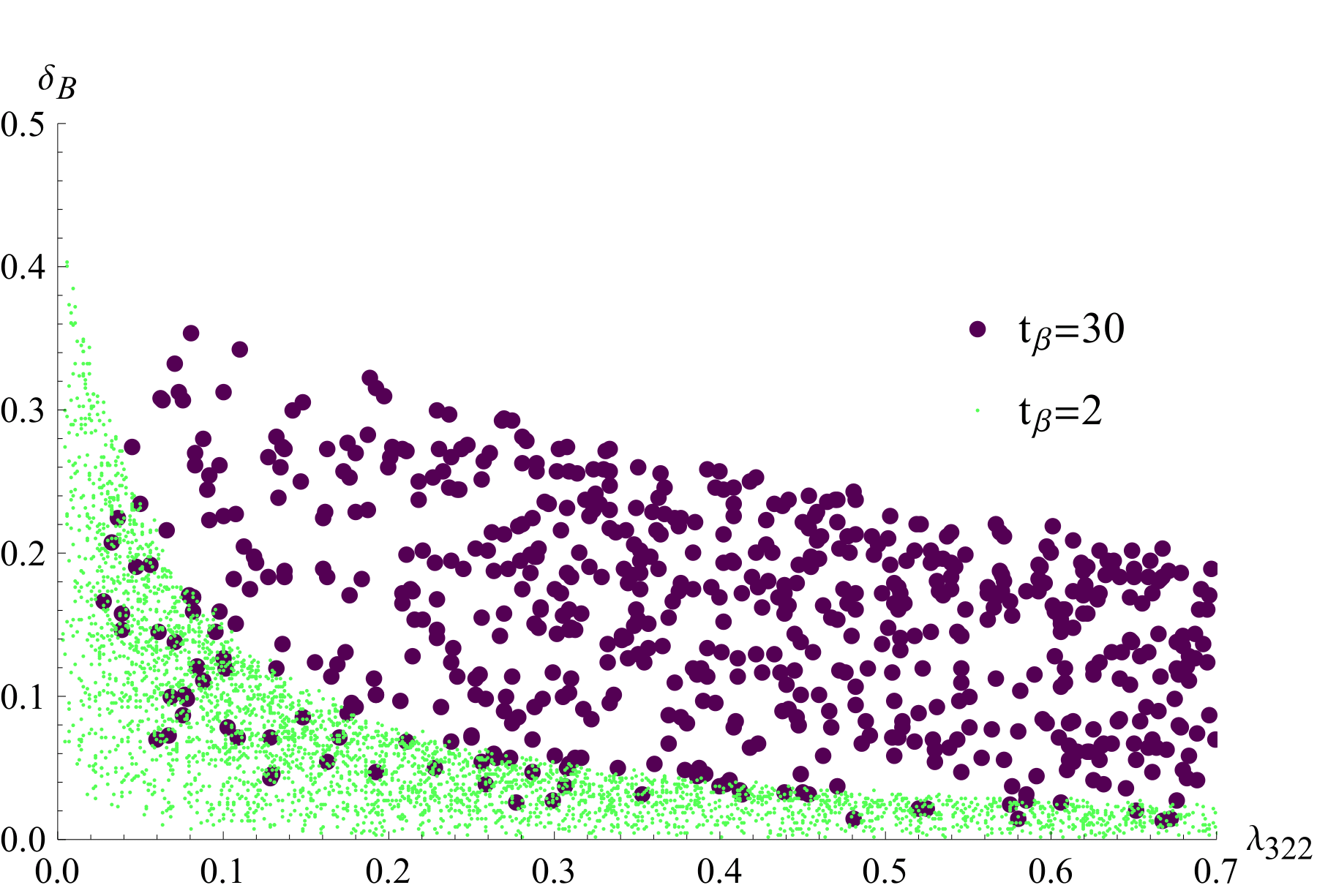}
  \caption{$\Delta\mu_{F\mu\mu}>0.8$}
  \label{fig:RPC0p8}
\end{subfigure}\\
\begin{subfigure}{.5\textwidth}
  \centering
  \includegraphics[width=.8\linewidth]{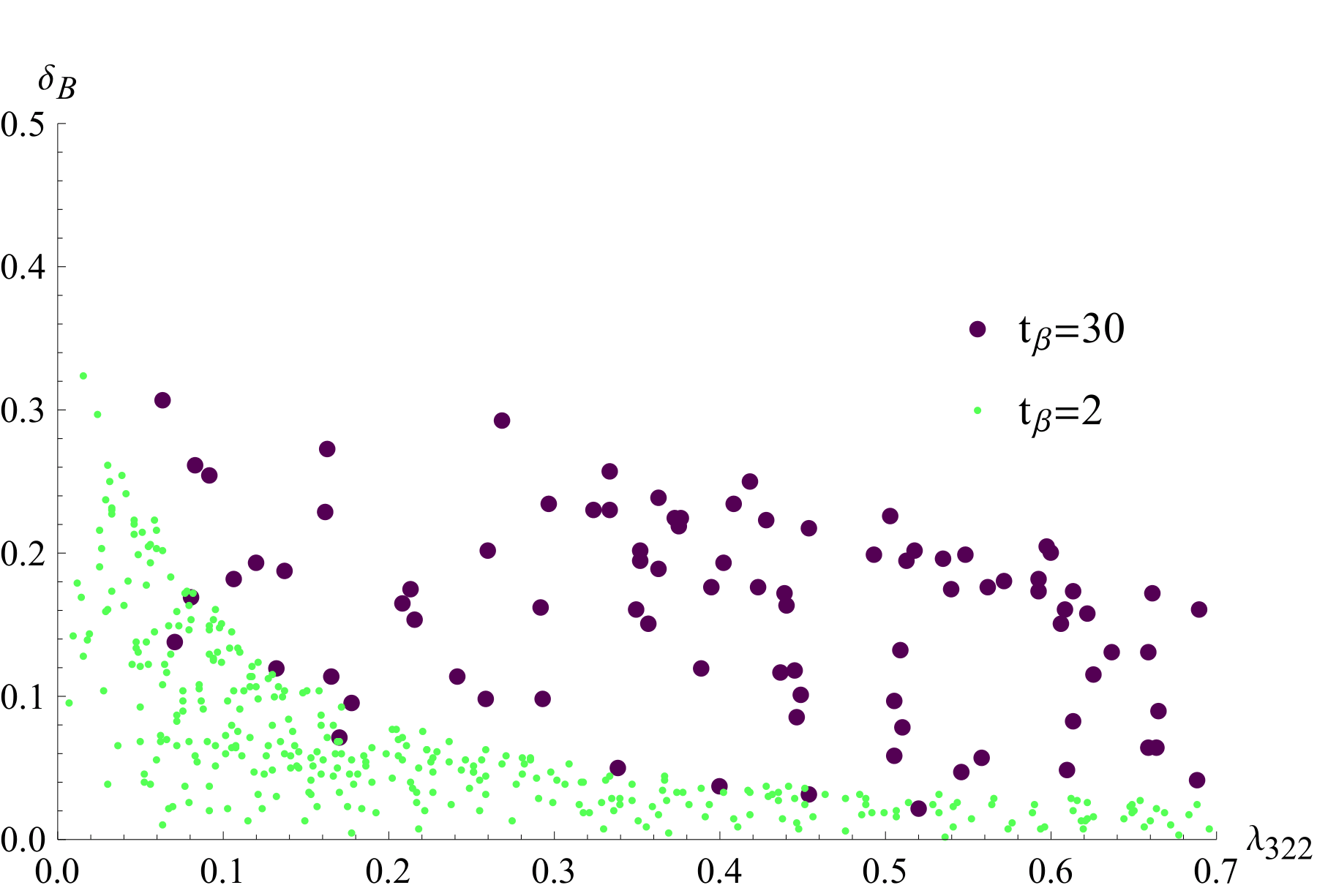}
  \caption{$\Delta\mu_{F\mu\mu}>0.95$}
  \label{fig:RPC0p95}
\end{subfigure}
\caption{Scatter-plots in the $\lambda_{322} - \delta_B$ RPV parameter plane, of RPV SUSY models that pass all the 
filters and constraints and yield 
$\Delta \mu_{F\mu\mu} > 0.5,~0.8$ and $0.95$, see 
text and eq.~\eqref{eq:RPCdistance}.}
\label{fig:TRPVscatterplots}
\end{figure}

In Figs.~\ref{fig:TRPVscatterplots}(a), (b) and (c) we give scatter plots in the $\lambda_{322} - \delta_B$ RPV parameter plane, corresponding to RPV SUSY models that pass all the 
filters and constraints and yield 
$\Delta \mu_{F\mu\mu} > 0.5,0.8,0.95$, respectively.    
We see for example, that a shift of up to ${\cal O}(100\%)$
in the $pp \to h \to \mu^+ \mu^-$ channel may be 
generated with values of the TRPV parameter 
$\lambda_{322} \sim {\cal O}(0.1)$, i.e., an order of magnitude smaller than its current upper 
bounds. 

\section{\label{sec:Summary}Summary}

We have explored the phenomenology of some variations of the RPV SUSY framework, 
confronting them with 
recent LHC data on the 125 GeV Higgs production and decay modes and with other
available constraints on the RPV parameter space.

We adopt a heavy SUSY scenario with TeV-scale squark 
and SU(2) singlet slepton masses as well as the decoupling 
limit in the SUSY Higgs sector, thereby considering multi-TeV masses for the 
heavy Higgs states. We then consider 
a simplified approach for both the Bilinear RPV (BRPV) and Trilinear RPV (TRPV) cases, by assuming non-negligible RPV effects only in a single generation, i.e., BRPV and TRPV interactions involving one sneutrino-flavor at a time, in most cases the 3rd generation sneutrino $\tilde\nu_\tau$. We show that the BRPV induced 
Higgs--sneutrino, lepton--gaugino and charged-Higgs--slepton mixings, give rise to new Higgs decay channels into lepton-gaugino pairs, with possible smoking gun 
RPV signatures of the form 
$h \to  \tau^\pm \ell^\mp + \missET$ ($\ell = e,\mu,\tau$), 
having rates several times larger than the expected SM (see eq.~\eqref{eq:BRPVdec2}) and/or RPC SUSY rates 
which are mediated by the Higgs decay $h \to WW^\star$. In some instances, when the SUSY spectrum contains an ${\cal O}(100)$ GeV light chargino, these signals are accompanied by an ${\cal O}(20-30\%)$ enhancement in the di-photon signal $pp \to h \to \gamma \gamma$. 
We also find that detectable BRPV effects of 
${\cal O}(20-30\%)$ might arise  
in some of the conventional Higgs signals, e.g., 
in $pp \to h \to \mu^+ \mu^-, \tau^+\tau^-$ which are unaffected by RPC SUSY effects in the decoupling limit and are, therefore, inherent to the RPV framework. 

We further examined TRPV scenarios and found that large 
RPV effects, in the range of $10-100\%$, can be 
generated in several Higgs production and decay modes, if the 125 GeV Higgs-like state is a Higgs-sneutrino BRPV mixed state and the 3rd or 2nd generation sneutrinos have 
${\cal O}(0.1-1)$ TRPV couplings to a pair of 
muons, $\tau$-leptons and/or to a pair of $d$ or $b$ quarks, i.e., $\tilde\nu_{\tau} \mu \mu$, 
$\tilde\nu_{\mu} \tau \tau$, $\tilde\nu_{\tau} dd$ or 
$\tilde\nu_{\tau} bb$. 
In particular, we find that detectable effects in the TRPV scenarios may arise in $pp \to h \to \mu^+ \mu^-, \tau^+\tau^-$ as well as in $pp \to Vh \to V b \bar b$ ($V=W,Z$).

We have provided specific benchmark models for the BRPV and TRPV scenarios 
and listed the corresponding SUSY parameter space and physical mass spectrum 
for all the above mentioned BRPV and TRPV effects. In Table~\ref{tab:RPVexpected} we list 
some of the notable RPV effects on the Higgs signals within these benchmark models.

\begin{table}
    \caption{\label{tab:RPVexpected}
   Expected RPV effects on the Higgs observables (signal strengths) within the benchmark RPV models considered in the paper. }
    \medskip{}
    \centering{}
\begin{tabular}{|c|c|c|c|}
\hline 
 & \multicolumn{3}{c|}{Production mode}\tabularnewline
\hline 
Decay Mode & $gg\to h$ & $hV$ & VBF\tabularnewline
\hline
\hline 
$h\to\gamma\gamma$ & $\mu_{F\gamma\gamma}^{\left(gg\right)}\sim1.26,\texttt{ BM}\lambda_{333}^{\prime}$ & - & $\mu_{V\gamma\gamma}^{\left(VBF\right)}\sim0.48,\texttt{ BM}\lambda_{311}^{\prime}$\tabularnewline
\hline 
$h\to ZZ^{\star}$ & SM-like & - & -\tabularnewline
\hline 
$h\to WW^{\star}$ & SM-like & - & -\tabularnewline
\hline 
$h\to b\overline{b}$ & - & $\mu_{Vbb}^{\left(hV\right)}\sim0.71,\texttt{ BM}\lambda_{311}^{\prime}$ & -\tabularnewline
\hline 
$h\to\tau^{+}\tau^{-}$ & $\mu_{F\tau\tau}^{\left(gg\right)}\sim\begin{cases}
0.73, & \texttt{BM}\tau\\
1.85, & \texttt{BM}\lambda_{233}\end{cases}$ & - & $\mu_{V\tau\tau}^{\left(VBF\right)}\sim\begin{cases}
0.65, & \texttt{BM}\lambda_{311}^{\prime}\\
1.85, & \texttt{BM}\lambda_{233}\end{cases}$\tabularnewline
\hline 
$h\to\mu^{+}\mu^{-}$ & $\mu_{F\mu\mu}^{\left(gg\right)}\sim\begin{cases}
0.75, & \texttt{BM}\mu\\
1.96, & \texttt{BM}\lambda_{322}\end{cases}$ & - & -\tabularnewline
\hline
\end{tabular}
\end{table}


\clearpage

\appendix

\section{\label{sec:Higgs Couplings}Higgs couplings, decays and production channels}

We list in this appendix part of the relevant analytical expressions 
for the couplings, decay channels and production mechanisms of the lightest CP-even 
Higgs-sneutrino mixed state, $h \equiv h_{RPV}$, in our BRPV 
SUSY framework. 
The complete set of Feynman rules for the BRPV and TRPV interactions can be found in 
\cite{SarahTRPVlink,Sarahlink}.  
In particular, following the notation of \cite{Sarahlink}, we focus below on direct BRPV effects, 
originating from F-terms, D-terms and from
interactions in the superpotential and soft breaking terms, highlighting the 
analytic features that arise in the BRPV scenario and are relevant to our work. 

\subsection{Higgs couplings and decays to heavy vector-bosons}

In the no-VEV BRPV basis, $v_{\tilde\nu_\tau}=0$, where the $\tau$-sneutrino doesn't 
condensate, the $hVV$ couplings 
($V=W^{\pm},Z$) are left unchanged with respect 
to the RPC case. That is, as in the RPC case, they scale as $s_{\beta-\alpha}=c_{\beta}Z_{h1}+s_{\beta}Z_{h2}$
relative to the corresponding SM coupling strength:
\begin{align}
\Lambda_{hVV} & =g_{V}^{SM}g_{hVV}^{RPC} ~, \label{eq:hVVcoupling}
\end{align}
where $g_{Z}^{SM}=\frac{1}{2}v\left(g_{1}s_{W}+g_{2}c_{W}\right)^{2}$,
$g_{W}^{SM}=\frac{1}{2}vg_{2}^{2}$, $g_{hVV}^{RPC}=c_{\beta}Z_{h1}+s_{\beta}Z_{h2}$ 
and the Higgs mixing elements $Z_{h1}$ and $Z_{h2}$ are determined by the 
diagonalization of the CP-even Higgs mass-squared matrix $m_{E}^2$ 
(see eq.~\eqref{eq:CP-even perturbed mass matrix}).

Thus, in our RPV setup the Higgs partial decay width to the vector-bosons as well as the $hV$ and VBF Higgs production channels (which are mediated by the $hVV$ coupling)
are also scaled by $g_{hVV}^{RPC}$ \cite{Decoupling}:
\begin{align}
\Gamma\left(h\to VV^\star \right) & =\left(g_{hVV}^{RPC}\right)^{2}\Gamma_{SM}\left(h\to VV^\star\right)\label{eq:hVVdecay} ~,
\end{align}
and
\begin{eqnarray}
\sigma\left( q \bar q \to V \to hV \right)  &=&
\left( g_{hVV}^{RPC} \right)^{2} 
\sigma_{SM} \left(q \bar q  \to V \to hV \right) ~, \\
\sigma\left(q q \to h qq \right)  &=& \left(g_{hVV}^{RPC}\right)^{2} \sigma_{SM}
\left(q q \to h qq  \right)
\label{eq:hVVprod} ~.
\end{eqnarray}

\subsection{Higgs couplings and decays to quarks and leptons}

The Higgs couplings to the quarks are also left unchanged with
respect to the RPC SUSY case, where they scale 
relative to the corresponding SM coupling strength as:
\begin{align}
\Lambda_{hq\overline{q}} & =g_{q}^{SM}g_{hq\overline{q}}^{RPC} ~, \label{eq:hqqcoupling}
\end{align}
with $g_{q}^{SM}=\frac{m_{q}}{v}$ and $g_{hu\overline{u}}^{RPC}=\frac{Z_{h2}}{s_{\beta}}$, 
$g_{hd\overline{d}}^{RPC}=\frac{Z_{h1}}{c_{\beta}}$, where $u(d)$ stands for an 
up(down)-quark. As in the RPC case, leptons which do not participate in the BRPV lepton-chargino mixing couple to the Higgs in similar fashion to the down-type quarks:
\begin{align}
\Lambda_{hll} & =g_{l}^{SM}g_{hll}^{RPC} ~, \label{eq:hllcoupling}
\end{align}
where $g_{l}^{SM}=\frac{m_{l}}{v}$ and $g_{hll}^{RPC}=\frac{Z_{h1}}{c_{\beta}}$. For the Higgs coupling to leptons participating in BRPV lepton-chargino mixing see Appendix~\ref{subsec:Higgs couplings and decays to gauginos}.

Thus, the Higgs partial decay width to the a pair of quarks is also scaled with respect to the SM \cite{Decoupling}:
\begin{align}
\Gamma\left(h\to q \bar q \right) & =\left(g_{hq \bar q}^{RPC}\right)^{2}\Gamma_{SM}\left(h\to q \bar q \right)\label{eq:hbbdecay} ~.
\end{align}
where
\begin{align}
\Gamma_{SM}\left(h\to q\overline{q}\right) & =N_{C}\frac{G_{F}m_{q}^{2}}{4\sqrt{2}\pi}m_{h}\left(1-\frac{4m_{q}^{2}}{m_{h}^{2}}\right)^{\frac{3}{2}}\label{eq:hbbdecaySM} ~.
\end{align}

Similar expressions also hold for the Higgs decay to RPC leptons by replacing $q\to l$ and setting $N_{C}=1$. The QCD corrections to eqs.~\eqref{eq:hbbdecaySM}--\eqref{eq:hbbdecay}
are important and were taken into account in our analysis, using the running masses evaluated at the scale of the Higgs mass, i.e., $m_{q}=\overline{m}_{q}\left(m_{h}\right)$
\cite{DjouadiSM,Decoupling}. In particular, for the 
$b$ and $c$ quarks we have $\overline{m}_{b}\left(m_{h}\right)\simeq2.8$
GeV and 
$\overline{m}_{c}\left(m_{h}\right)\simeq0.6$ GeV, respectively.

\subsection{\label{subsec:Higgs couplings to squarks and sleptons}Higgs couplings to squarks and sleptons}

The couplings of the lightest CP-even Higgs-sneutrino mixed state to the squarks
and sleptons 
are relevant in this work for their contributions to 
the 1-loop decays $h \to gg$ and $h \to \gamma \gamma$ and in the calculation 
of the higher-order corrections to the Higgs mass.  
We note that the contribution of D-terms to the squark mass matrices is negligible for 
multi-TeV squarks, as assumed throughout this work.

In the BRPV scenario within the no-VEV basis, 
the left-right mixing matrices of the up and
down-type squarks, $Z^{U}$ and $Z^{D}$, respectively \cite{Sarahlink}, 
remain unchanged with respect to the RPC case. 
In particular, the Higgs couplings to the down-type squarks 
are 
equal to their values in the RPC case. 
In particular, there are no new F-terms due to BRPV in the down-squark sector and the BRPV D-terms vanish in the no-VEV basis.
Thus, the 
$h \tilde{b}_i \tilde{b}_j$ coupling is ($\tilde{b}$ is a bottom-squark):\footnote{We note that there is a TRPV F-term  
in the $h \tilde b \tilde b$ coupling which is $\propto y_b \lambda^{'}_{333} Z_{h3}$. This term indirectly affects the 1-loop $h gg$ and $h\gamma\gamma$ vertices, but it is negligible for our purpose mainly due to the $1/m_{\tilde{b}}^2$ suppression in these 1-loop couplings.}

\begin{align}
\Lambda_{h\tilde{b_{i}}\tilde{b}_{j}} & =\frac{g_{2}}{m_{W}}g_{h\tilde{b_{i}}\tilde{b}_{j}}^{RPC} ~, \label{eq:hb_b_coupling}
\end{align}
where we have defined the "reduced" RPC coupling $g_{h\tilde{b_{i}}\tilde{b}_{j}}^{RPC}$ 
and factored out 
the term $g_{2}/m_{W}=2/v$ for later use (see e.g., \cite{Decoupling}).
The full expression for $\Lambda_{h\tilde{b_{i}}\tilde{b}_{j}}$ can be found
in \cite{Sarahlink}.

On the other hand, the Higgs couplings to a pair of up-type squarks do receive a new BRPV 
F-term contribution which is 
$\propto y_{u} \mu \delta_{\epsilon} Z_{h3}$ (recall that 
$Z_{h3}=Z_{h3}(\delta_B)$). 
In particular, for the top-squarks the BRPV F-term can be significant and we have (see also \cite{Sarahlink}): 
\begin{align}
\Lambda_{h\tilde{t_{i}}\tilde{t_{j}}} & =\frac{g_{2}}{m_{W}}\left(g_{h\tilde{t}_{i}\tilde{t}_{j}}^{RPC}-\frac{m_{t}}{s_{\beta}}Z_{i1}^{U}Z_{j2}^{U}\mu 
\delta_{\epsilon} Z_{h3}\right) ~, \label{eq:ht_t_coupling}
\end{align}
where again
we factored out the term $g_{2}/m_{W}=2/v$ and introduced  
the "reduced" RPC coupling $g_{h\tilde{t}_{i}\tilde{t}_{j}}^{RPC}$.
Note that since $Z_{h3}$
depends on the soft BRPV term $\delta_{B}$, the new BRPV
term in eq.~\eqref{eq:ht_t_coupling}
contains two BRPV insertions, 
i.e., $\delta_{\epsilon} \times \delta_{B}$. 
It also modifies the contribution of the top-squark loop in the $ggh$ vertex, thereby 
changing the gluon-fusion Higgs production mode; this effect is taken into account in our analysis. 

The BRPV couplings $\delta_\epsilon$ and $\delta_B$ also generate mixing between the charged Higgs states and the charged sleptons. In particular, 
assuming only a 3rd generation BRPV scenario, the slepton--charged Higgs mass matrix
in the $\left(H_{d}^{-},H_{u}^{+},\tilde{\tau}_{L},\tilde{\tau}_{R}\right)$
weak basis reads \cite{Sarahlink}:\footnote{In some instances we apply the single generation BRPV working assumption to the 2nd generation, in which case the slepton--charged Higgs mass matrix can be similarly written in the $\left(H_{d}^{-},H_{u}^{+},\tilde{\mu}_{L},\tilde{\mu}_{R}\right)$
weak basis and the change in the index $\tau \to \mu$ should be applied in eq.~\eqref{eq:sleptonmassmatrix} in all the relevant entries.}
\begin{equation}
m_{\tilde{\tau}}^{2} =
\begin{psmallmatrix}
    m_{W}^{2}s_{\beta}^{2}+m_{A}^{2}s_{\beta}^{2} & m_{W}^{2}s_{\beta}c_{\beta}+\frac{1}{2}m_{A}^{2}s_{2\beta} & -\delta_{B}s_{\beta}^{2}m_{A}^{2} & -\delta_{\epsilon}\mu m_{\tau}t_{\beta}\\
    m_{W}^{2}s_{\beta}c_{\beta}+\frac{1}{2}m_{A}^{2}s_{2\beta} & m_{W}^{2}c_{\beta}^{2}+m_{A}^{2}c_{\beta}^{2} & -\frac{1}{2}\delta_{B}m_{A}^{2}s_{2\beta} & -\delta_{\epsilon}\mu m_{\tau}\\
    -\delta_{B}s_{\beta}^{2}m_{A}^{2} & -\frac{1}{2}\delta_{B}m_{A}^{2}s_{2\beta} & m_{\tau}^{2}+m_{\tilde{\nu}_{\tau}}^{2}-m_{W}^{2}\left(c_{\beta}^{2}-s_{\beta}^{2}\right) & \left(A_{\tau}-\mu t_{\beta}\right)m_{\tau}\\
    -\delta_{\epsilon}\mu m_{\tau}t_{\beta} & -\delta_{\epsilon}\mu m_{\tau} & \left(A_{\tau}-\mu t_{\beta}\right)m_{\tau} & m_{\tau}^{2}+m_{\tilde{\tau}_{RR}}^{2}-\frac{1}{4}g_{1}^{2}v\left(c_{\beta}^{2}-s_{\beta}^{2}\right) ~,
\end{psmallmatrix}
\label{eq:sleptonmassmatrix}
\end{equation}
where $m_{\tilde{\nu}_{\tau}}^{2}$ is defined in eq.~\eqref{eq:msndependencies},
$m_{\tilde{\tau}_{RR}}^{2}$ is the right-handed soft mass of the 3rd generation slepton, 
$\tilde\tau$, and we have used the minimization conditions and definitions in eqs.~\eqref{eq:minimaconditionI}--\eqref{eq:eps_redefinition}. Also, we have used 
the MFV relation $A_{\tau} \propto y_{\tau}$ by 
generically setting  
$A_f \equiv y_f \cdot \tilde A$.
Note that, as opposed to the squark sector, in the mass matrix $m_{\tilde{\tau}}^{2}$ of eq.~\eqref{eq:sleptonmassmatrix}
we have kept the D-terms, since their relative effect is larger in the slepton sector.

The weak states $\left(H_{d}^{-},H_{u}^{+},\tilde{\tau}_{L},\tilde{\tau}_{R}\right)$
are given in terms of the physical states $\left(\tilde{\tau}_{j}\right)$
by 
\begin{subequations}
\begin{align}
H_{d}^{-} & =Z_{j1}^{+}\tilde{\tau}_{j}\label{eq:Hdminustau}\\
H_{u}^{+} & =Z_{j2}^{+}\tilde{\tau}_{j}\label{eq:Huplustau}\\
\tilde{\tau}_{L} & =Z_{j3}^{+}\tilde{\tau}_{j}\label{eq:stauleftstau}\\
\tilde{\tau}_{R} & =Z_{j4}^{+}\tilde{\tau}_{j}\label{eq:staurightstau}
\end{align}
\end{subequations}
where $\tilde{\tau}_{j}$ corresponds to the massless Goldstone boson and $\tilde{\tau}_{2,3,4}$ are the 
physical states which are added in our analysis (e.g., in the 1-loop decay $h \to \gamma \gamma$) 
although their effect on the 125 GeV Higgs physics
is small in general in the decoupling limit \cite{Decoupling}.

As mentioned above, the Higgs couplings to the charged sleptons--charged Higgs mixed states are needed for the calculation of the 
1-loop $h\rightarrow\gamma\gamma$ 
decay and for the higher-order corrections to the Higgs mass. These quantities require 
only the diagonal $h\tilde{\tau_{i}}\tilde{\tau_{i}}$ couplings which are given by:
\begin{align}
\Lambda_{h\tilde{\tau_{i}}\tilde{\tau_{i}}} & =\frac{g_{2}}{m_{W}}\Bigl[g_{h\tilde{\tau}_{i}\tilde{\tau}_{i}}^{RPC}-v^{2}c_{\beta}\frac{g_{2}^{2}}{4}Z_{i1}^{+}Z_{i3}^{+}Z_{h3}+\frac{m_{\tau}}{c_{\beta}}A_{\tau}Z_{i1}^{+}Z_{i4}^{+}Z_{h3}+\frac{m_{\tau}^{2}}{c_{\beta}}Z_{i1}^{+}Z_{i3}^{+}Z_{h3}+\nonumber \\
 & +\delta_{\epsilon}\mu\frac{m_{\tau}}{c_{\beta}}Z_{i1}^{+}Z_{i4}^{+}Z_{h2}-v^{2}s_{\beta}\frac{g_{2}^{2}}{4}Z_{h3}Z_{i3}^{+}Z_{i2}^{+}+\mu\frac{m_{\tau}}{c_{\beta}}Z_{i4}^{+}Z_{h3}Z_{i2}^{+}+\delta_{\epsilon}\mu\frac{m_{\tau}}{c_{\beta}}Z_{i4}^{+}Z_{h1}Z_{i2}^{+}\Bigr]\label{eq:hstaustaucoupling} ~,
 \end{align}
where the term $g_{2}/m_{W}=2/v$ is again factored out. 
We can see from eq.~\eqref{eq:hstaustaucoupling}
that the BRPV D-terms ($\propto g_{2}^{2})$ correspond 
to the RPC sneutrino--slepton--charged Higgs and sneutrino--slepton--slepton
couplings, and thus depend on the BRPV mixing parameter $\delta_B$ through the $Z_{h3}$ rotation. 
Also, the $A_{\tau}$
term in eq.~\eqref{eq:hstaustaucoupling} originates from the RPC
sneutrino--slepton--charged Higgs 
trilinear coupling. The rest of the
terms in eq.~\eqref{eq:hstaustaucoupling} are new BRPV F-terms; the ones that involve the RPC sneutrino depend on $\delta_{B}$ (i.e., through $Z_{h3}$), while the others are proportional to $\delta_{\epsilon}$. 

\bigskip

\subsection{\label{subsec:Higgs couplings and decays to gauginos}Higgs couplings to Gauginos}

The Higgs couplings to the gauginos can be written 
in a general form as \cite{Sarahlink}:
\begin{eqnarray}
\Lambda_{h \tilde\chi_{i}^0 \tilde\chi_{j}^0/h \chi_{i}^+ \chi_{j}^-} = 
 \Lambda_{Lij}^{N/C} L  +\Lambda_{Rij}^{N/C} R  ~,
\label{eq:gaugino-coup}
\end{eqnarray}
where $R(L)=(1\pm\gamma_5)/2$ and the left and right-handed couplings, $\Lambda_{L/R ij}^{N/C}$ 
depend on the BRPV parameters 
$\delta_{\epsilon}$ and $\delta_B$. In particular, 
we have:
\begin{eqnarray}
\Lambda_{R/Lij}^{N/C} \equiv g_{R/Lij}^{N/C \left(\delta_{\epsilon}\right)}+
g_{R/Lij}^{N/C \left(\delta_{\epsilon}, \delta_{B}\right)} ~,\label{eq:hinvisiblecoupling}
\end{eqnarray}
where $g_{R/Lij}^{N/C \left(\delta_{\epsilon}\right)}$ depend on $\delta_\epsilon$ and on the 
elements $Z_{h1}$ and $Z_{h2}$ (which are independent 
of $\delta_{B}$), while 
$g_{R/Lij}^{N/C \left(\delta_{\epsilon}, \delta_{B}\right)}$
are proportional to the Higgs-snuetrino mixing element $Z_{h3}$ which contain a $\delta_{B}$ insertion. 
The couplings $g_{R/Lij}^{N/C \left(\delta_{\epsilon}, \delta_{B}\right)}$ vanish as $\delta_B \to 0$. Their 
explicit 
form is: 
\begin{eqnarray}
{\rm Neutralinos:} &~~& g_{Lij}^{N \left(\delta_{\epsilon},\delta_{B}\right)} =g_{Rij}^{N \left(\delta_{\epsilon},\delta_{B}\right)}=
\frac{1}{2}\left(g_{1}U_{N_{j2}}U_{N_{i1}}-g_{2}U_{N_{j3}}U_{N_{i1}}+i\leftrightarrow j\right)Z_{h3} ~, \label{eq:gLRepsdeltahinvisible} \\
{\rm Charginos:} &~~& g_{Lij}^{C \left(\delta_{\epsilon},\delta_{B}\right)} = 
\left( \frac{e}{\sqrt{2} s_W} U_{R_{i2}}U_{L_{j1}} - \frac{m_\tau}{v c_\beta} U_{R_{i1}}U_{L_{j3}} \right) Z_{h3} ~, \label{eq:gLRcharginos1} \\
&~~& g_{Rij}^{C \left(\delta_{\epsilon},\delta_{B}\right)} = 
\left( \frac{e}{\sqrt{2} s_W} U_{L_{i1}}U_{R_{j2}} - \frac{m_\tau}{v c_\beta} U_{L_{i3}}U_{R_{j1}} \right) Z_{h3} ~,
\label{eq:gLRcharginos2}
\end{eqnarray}
where $U_N$ is the neutralino mixing matrix (see eq.~\eqref{eq:UN}) and $U_{L,R}$ 
are the chargino mixing matrices (see eq.~\eqref{eq:SVD_chargino}). 
The explicit form of the couplings $g_{R/Lij}^{N/C \left(\delta_{\epsilon}\right)}$
are not very enlightening and will not be given here.

In terms of the above couplings, the widths for the decays 
$h \to \tilde\chi^0_i \tilde\chi^0_j$ and 
$h \to \chi^{+}_i \chi^{-}_j$
are given by: 
\begin{eqnarray}
\Gamma\left(h\to\tilde{\chi}_{i}^{0}{\tilde\chi}_{j}^{0}/\chi^{+}_i \chi^{-}_j\right) &=& \left[\left(\left|\Lambda_{Lij}^{N/C}\right|^{2}+\left|\Lambda_{Rij}^{N/C}\right|^{2}\right)\left(m_{h}^{2}-
m_{\tilde{\chi}_{i}^{0}/\chi_{i}^{+}}^{2}-
m_{\tilde{\chi}_{j}^{0}/\chi_{j}^{+}}^{2}\right) \right. \nonumber \\
&-& \left. 4\;\text{Re}\left\{ \Lambda_{Lij}^{N/C}\Lambda_{Rij}^{N/C}\right\} m_{\tilde{\chi}_{i}^{0}/\chi_{i}^{+}}
m_{\tilde{\chi}_{j}^{0}/\chi_{j}^{+}}
\right] \times\frac{\lambda^{\frac{1}{2}}\left(m_{h}^{2},
 m_{\tilde{\chi}_{i}^{0}/\chi_{i}^{+}}^{2},
 m_{\tilde{\chi}_{j}^{0}/\chi_{j}^{+}}^{2}\right)}{16\pi m_{h}^{3}}\label{eq:hinvdecay} ~, \nonumber \\
\end{eqnarray}
where $i,j=1-5$ for neutralinos and $i,j=1-3$ for the charginos, assuming a single generation BRPV mixing in both sectors. Also, 
$\lambda\left(x,y,z\right)=\left(x-y-z\right)^{2}-4yz$ and the gaugino couplings $\Lambda_{L,Rij}^{N/C}$ are defined in eq.~\eqref{eq:gaugino-coup} (their full expressions are given in 
\cite{Sarahlink}). 

\begin{figure}[htb]
    \begin{centering}
    \includegraphics[width=0.80\textwidth]{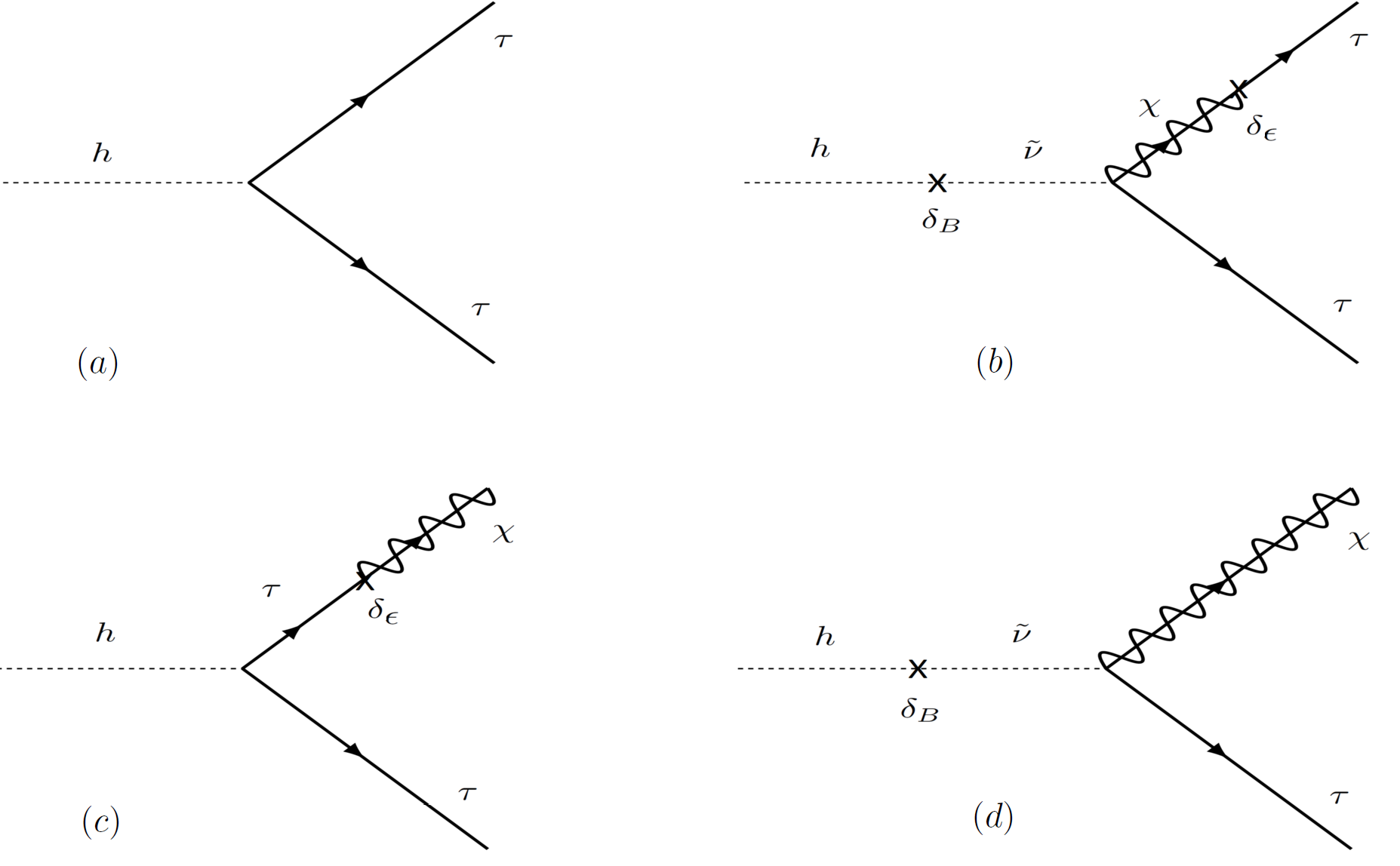}
    \par\end{centering}
    \caption{\label{fig:hchrydiags}
        Sample diagrams of the Higgs couplings/decays $h\rightarrow \tau^+\tau^-$
        (diagrams (a) and (b)) and $h \to \tau^\pm \chi^\mp$ (diagrams (c) and (d)).}
\end{figure}

\begin{figure}[htb]
    \begin{centering}
    \includegraphics[width=0.65\textwidth]{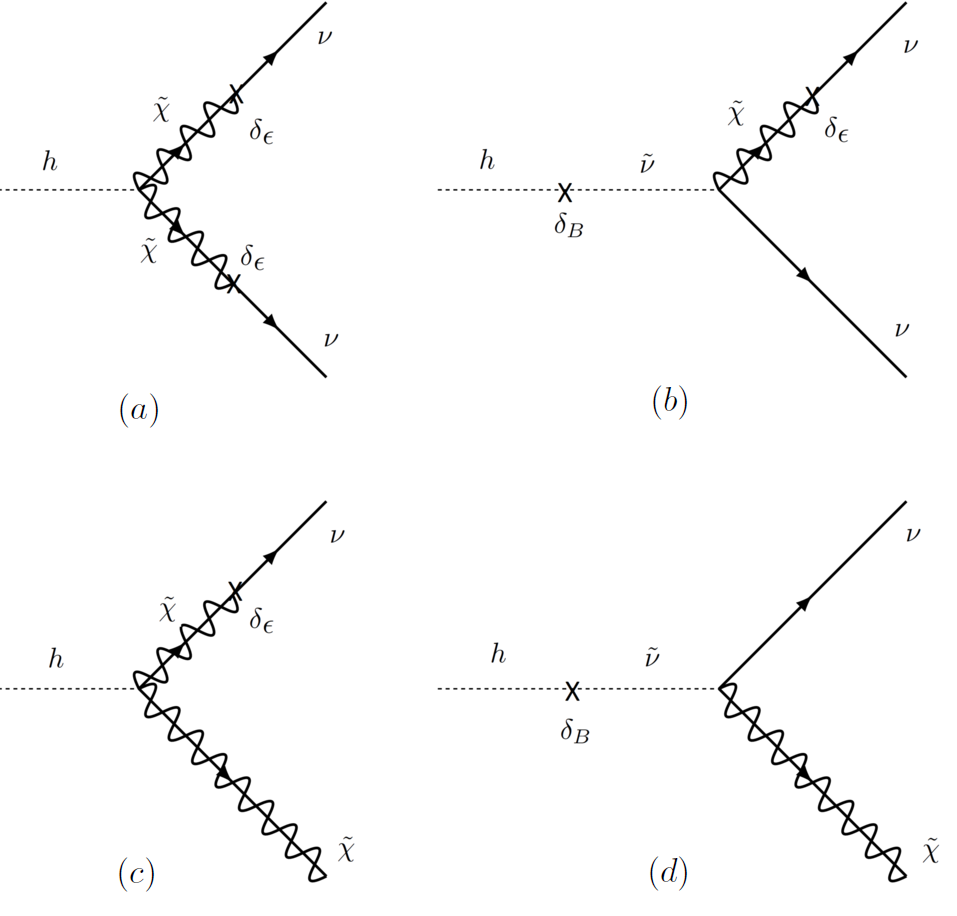}
    \par\end{centering}
    \caption{\label{fig:hinvdiags} 
    Diagrammatic description of the Higgs couplings/decays $h\rightarrow \nu_\tau \nu_\tau$
        (diagrams (a) and (b)) and $h \to \nu_\tau \tilde\chi^0$ (diagrams (c) and (d)).}
\end{figure}

We identify the lightest neutralino (RPV) state $\tilde\chi^0_1$ as
the $\tau$-neutrino, $\nu_{\tau} \equiv \tilde\chi^0_1$,
and the lightest chargino as the $\tau$-lepton, $\tau^+ \equiv \chi^{+}_{1}$, and we focus  
in section~\ref{subsec:Higgs decays to Gauginos} 
on the Higgs decays 
$h \to \nu_\tau \nu_\tau, \nu_\tau \tilde\chi_{2}^{0}$ 
and $h \to \tau^+ \tau^-, \tau^\pm \chi_{2}^\mp$, which 
corresponds to 
$h \to \tilde\chi_{1}^{0} \tilde\chi_{1}^{0}, \tilde\chi_{1}^{0} \tilde\chi_{2}^{0}$ and $h \to \chi_{1}^{+} \chi_{1}^{-}, \chi_{1}^{\pm} \chi_{2}^{\mp}$, respectively.
Sample diagrams of the couplings which generate these Higgs decays to a 
lepton-gaugino pair are given in Figs.~\ref{fig:hchrydiags} and \ref{fig:hinvdiags}. 

We also note the following:
\begin{itemize}
    \item The $h\tau^+ \chi_2^-$ as well as the $h \nu_\tau \nu_\tau$ (for $m_{\nu_\tau} \to 0$) 
    and $h \nu_\tau \tilde\chi_2^0$
    couplings have no RPC equivalent and are, therefore, 
    pure RPV couplings.
    \item The $h\tau^+ \chi_2^-$ and
     $h \nu_\tau \tilde\chi_2^0$ RPV couplings have a term proportional only to $Z_{h3}=Z_{h3}(\delta_B)$ (see  eqs.~\eqref{eq:gLRepsdeltahinvisible} and \eqref{eq:gLRcharginos2}). These terms are new BRPV
     D-terms which are generated from the RPC sneutrino--Wino/Higgsino--$\tau$ and sneutrino--Zino--neutrino interactions, respectively, due to the $\tilde{\nu}_{\tau} - h$ mixing effect.
     \item Both $h\tau^+ \chi_2^-$ and
     $h \nu_\tau \tilde\chi_2^0$ RPV couplings also have a pure BRPV contribution from the superpotential,   
which depend only on $\delta_{\epsilon}$ (see diagrams
Fig.~\ref{fig:hchrydiags}(c) and Fig.~\ref{fig:hinvdiags}(c)). 
\item The RPV coupling of a Higgs to a pair of $\tau$-neutrinos $\Lambda_{h\nu_\tau \nu_\tau}$ (which in the RPC limit vanish for $m_{\nu_\tau} \to 0$) contains two BRPV insertions, being proportional to either $\delta_\epsilon^2$ or 
to $\delta_\epsilon \cdot \delta_B$. This coupling is, therefore, suppressed with respect to 
$\Lambda_{h \nu_\tau \tilde\chi_{2}^0}$, which as mentioned above, 
can be generated with a single BRPV insertion. Indeed, 
this is verified in our numerical
simulations where we find that $h\to \nu_\tau \nu_\tau$
is suppressed by several orders of magnitude compared to $h \to \nu_\tau \tilde{\chi}_{2}^{0}$ (i.e., when $m_{\tilde{\chi}_{2}^{0}}<m_{h}$).
\end{itemize}

\subsection{\label{subsec:1-loop decay to photons}The 1-loop decay $h\to\gamma\gamma$}

The Higgs decay to a pair of photons in the SM is given by \cite{DjouadiSM}:
\begin{align}
\Gamma_{SM}\left(h\to\gamma\gamma\right) & =\frac{G_{F}\alpha^{2}m_{h}^{3}}{128\sqrt{2}\pi^{3}}\biggl|\sum_{f}N_{C}Q_{f}^{2}A_{\frac{1}{2}}\left(\tau_{f}\right)+A_{1}\left(\tau_{W}\right)\biggr|^{2}\label{eq:hphotondecaySM}
\end{align}
where $\tau_{i}=\frac{4m_{i}^{2}}{m_{h}^{2}}$ and the expressions for the loop functions $A_{\frac{1}{2}}$ (for a fermion loop) and $A_{1}$ (for the 
$W$ loop) can be found in \cite{DjouadiSM}. The dominant SM contributions arise from the top-quark and W-boson loop exchanges. 

In the BRPV SUSY framework, the Higgs decay to a pair of photons can be cast in the following form \cite{Decoupling}:
\begin{align}
\Gamma\left(h\to\gamma\gamma\right) = \nonumber \\
\frac{G_{F}\alpha^{2}m_{h}^{3}}{128\sqrt{2}\pi^{3}} & \times\Biggl|\sum_{q=t,b}N_{C}Q_{f}^{2}g_{hq\overline{q}}^{RPC}A_{\frac{1}{2}}\left(\tau_{q}\right)+g_{hVV}^{RPC}A_{1}\left(\tau_{W}\right)\nonumber \\
 & +\sum_{i=1}^{3} \frac{2m_{W}}{m_{\chi_{i}^{\pm}}} \frac{\Lambda^C_{ii}}{e} A_{\frac{1}{2}}\left(\tau_{\chi_{i}^{\pm}}\right)\nonumber \\
 & + \frac{v}{2} \sum_{i=1}^{2}\left[\frac{\Lambda_{h\tilde{b}_{i}\tilde{b}_{i}}}{m_{\tilde{b}_{i}}^{2}}N_{C}Q_{\tilde{b}}^{2}A_{0}\left(\tau_{\tilde{b}_{i}}\right)+
 \frac{\Lambda_{h\tilde{t}_{i}\tilde{t}_{i}}}{m_{\tilde{t}_{i}}^{2}}N_{C}Q_{\tilde{t}}^{2}A_{0} \left(\tau_{\tilde{t}_{i}}\right)\right]\nonumber \\
 & +\frac{v}{2} \sum_{i=2}^{4}\frac{\Lambda_{h\tilde{\tau_{i}}\tilde{\tau_{i}}}}{m_{\tilde{\tau}_{i}}^{2}}A_{0}\left(\tau_{\tilde{\tau}_{i}}\right)\Biggr|^{2}
\label{eq:hphotondecay}
\end{align}
where $\Lambda_{ii}^C = \Lambda_{Lii}^C=\Lambda_{Rii}^C$ and 
$\Lambda_{h \tilde{f}_i \tilde{f}_i}$ ($f=\tau,b,t$) 
are the diagonal Higgs couplings to the charginos and sfermions which are defined above. 

We note that, for the heavy sfermion spectrum that is considered in this work, the chargino contributions to 
$h \to \gamma \gamma$ are much larger than the sfermions one.

\subsection{The 1-loop decay $h\to gg$}

In the SM, the loop-induced Higgs decay to a pair of gluons 
is given by \cite{DjouadiSM}:
\begin{align}
\Gamma_{SM}\left(h\to gg\right) & =K_{SM}^{QCD}\frac{G_{F}\alpha_{s}^{2}m_{h}^{3}}{36\sqrt{2}\pi^{3}}\left|\frac{3}{4}\sum_{q=t,b}A_{\frac{1}{2}}\left(\tau_{q}\right)\right|^{2}\label{eq:gghproductionSM} ~,
\end{align}
where the QCD corrections
to eq.~\eqref{eq:gghproductionSM} are taken into
account by the QCD K-factor and by using the quark running masses evaluated at the scale of the Higgs mass: $m_{q}=\overline{m}_{q}\left(m_{h}\right)$
\cite{DjouadiSM}. In particular, the SM QCD K-factor 
is $K_{SM}^{QCD}\simeq1.84 $ and $\overline{m}_{b}\left(m_{h}\right)\simeq2.8$
GeV, $\overline{m}_{c}\left(m_{h}\right)\simeq0.6$ GeV. 

In the  BRPV SUSY framework, the loop-induced 
$h \to gg$ amplitude receives additional contributions from the squarks; the dominant ones are generated by the sbottoms and stops \cite{Decoupling}: 

\begin{align}
\Gamma\left(h\to gg\right)=K^{QCD}\frac{G_{F}\alpha_{s}^{2}m_{h}^{3}}{36\sqrt{2}\pi^{3}} & \times\biggl|\frac{3}{4}\sum_{q=t,b}g_{hq\overline{q}}^{RPC}A_{\frac{1}{2}}\left(\tau_{q}\right)\nonumber \\
 & +\frac{3v}{8} \sum_{i=1}^{2} \left[ \frac{\Lambda_{h\tilde{b}_{i}\tilde{b}_{i}}}{m_{\tilde{b}_{i}}^{2}}A_{0}\left(\tau_{\tilde{b}_{i}}\right) + 
 \frac{\Lambda_{h\tilde{t}_{i}\tilde{t}_{i}}}{m_{\tilde{t}_{i}}^{2}}A_{0}\left(\tau_{\tilde{t}_{i}}\right)
 \right] \biggl|^{2}\label{eq:gghproduction} ~.
 \end{align} 

In the decoupling limit the squarks contribution is subdominant and, since in this limit we also have to a
good approximation 
$K^{QCD}\simeq K_{SM}^{QCD}$ \cite{Decoupling}, we expect a small deviation in $\Gamma(h \to gg)$ compared to the corresponding RPC SUSY rate and, therefore, also a small deviation in the gluon-fusion Higgs 
production mechanism. 

\newpage

\section{\label{sec:SUSY spectra}RPV SUSY Spectra of the benchmark models}

In this appendix we list the SUSY spectrum for each of the benchmark models in sections~\ref{sec:Numerical-results}--\ref{sec:TRPV}.
The SUSY spectrum of the BRPV benchmark models \texttt{BM1A}-\texttt{BM1B}, \texttt{BM2}-\texttt{BM3} and \texttt{BM$\mu$}-\texttt{BM$\tau$} are given in Tables~\ref{tab:SUSY-spectrum-BRPV1}, \ref{tab:SUSY-spectrum-BRPV2} and \ref{tab:SUSY-spectrum-BRPV3}, respectively, and the SUSY spectrum of the TRPV benchmark models \texttt{BM$\lambda^{\prime}_{333}$}-\texttt{BM$\lambda^{\prime}_{311}$} and \texttt{BM$\lambda_{233}$}-\texttt{BM$\lambda_{322}$} are given in Tables~\ref{tab:SUSY-spectrum-TRPV1} and \ref{tab:SUSY-spectrum-TRPV2}, respectively. 
\begin{table}[htb]
    \caption{\label{tab:SUSY-spectrum-BRPV1}
    SUSY mass spectrum corresponding to the
    benchmark models \texttt{BM1A} and \texttt{BM1B}. We have denoted by $\tilde \chi_{i}^{0}$ the neutralino states, by $\chi_{i}^{\pm}$ the chargino states,  $(\Tilde{\nu}_{\tau}^{E},H)$ are the CP-even sneutrino and heavy Higgs states, respectively and $(\Tilde{\nu}_{\tau}^{O},A)$ are the CP-odd sneutrino and pseudoscalar Higgs, respectively. Also, $\Tilde{\tau}_{i}$ are the slepton states, $\Tilde{t}_{i}$ are the stops and $\Tilde{b}_{i}$ are the sbottom states. All masses are given in GeV.}
    \medskip{}
    \centering{}
    \begin{tabular}{ccccc}
        \toprule 
         & \texttt{BM1A} & \texttt{BM1B}  \tabularnewline
        \midrule
        \addlinespace
        $(\tilde\chi_{2}^{0},\tilde\chi_{3}^{0},\tilde\chi_{4}^{0},\tilde\chi_{5}^{0})$ & $(94.2,510.4,628.3,650.7)$ & $(90.1,97.6,1034.6,2031.4)$ \tabularnewline
        \addlinespace
        $(\chi_{2}^{\pm},\chi_{3}^{\pm})$ & $(94.3,638.7)$ & $(91.9,1034.6)$  \tabularnewline
        \addlinespace
        $(\Tilde{\nu}_{\tau}^{E},H)$ & $(200.2,4473.6)$ & $(162.8,2574.3)$  \tabularnewline
        \addlinespace
        $(\Tilde{\nu}_{\tau}^{O},A)$ & $(200.1,4472.8)$ & $(162.4,2573.4)$  \tabularnewline
        \addlinespace
        $(\Tilde{\tau}_{2},\Tilde{\tau}_{3},\Tilde{\tau}_{4})$ & $(210.2,1509.2,4473.5)$ & $(176.8,1303.9,2574.6)$ \tabularnewline
        \addlinespace
        $(\Tilde{t}_{1},\Tilde{t}_{2})$ & $(6056.3,6091.8)$ & $(2779.5,2949.3)$  \tabularnewline
        \addlinespace
        $(\Tilde{b}_{1},\Tilde{b}_{2})$ & $(4814.4,6071.6)$ & $(2860.5,4996.4)$ \tabularnewline
        \bottomrule
        \addlinespace
    \end{tabular}
\end{table}
\begin{table}[htb]
    \caption{\label{tab:SUSY-spectrum-BRPV2}
    SUSY mass spectrum corresponding to the
    benchmark models \texttt{BM2} and \texttt{BM3}. See also caption to Table~\ref{tab:SUSY-spectrum-BRPV1}.}
    \medskip{}
    \centering{}
    \begin{tabular}{ccccc}
        \toprule 
         & \texttt{BM2} & \texttt{BM3}  \tabularnewline
        \midrule
        \addlinespace
        $(\tilde\chi_{2}^{0},\tilde\chi_{3}^{0},\tilde\chi_{4}^{0},\tilde\chi_{5}^{0})$ & $(93.5,223.7,227.6,999.2)$ & $(86.0,125.9,163.6,1006.2)$ \tabularnewline
        \addlinespace
        $(\chi_{2}^{\pm},\chi_{3}^{\pm})$ & $(215.9,999.2)$ & $(118.6,1006.3)$  \tabularnewline
        \addlinespace
        $(\Tilde{\nu}_{\tau}^{E},H)$ & $(229.8,2748.4)$ & $(163.2,3179.2)$  \tabularnewline
        \addlinespace
        $(\Tilde{\nu}_{\tau}^{O},A)$ & $(229.5,2747.6)$ & $(163.0,3178.7)$  \tabularnewline
        \addlinespace
        $(\Tilde{\tau}_{2},\Tilde{\tau}_{3},\Tilde{\tau}_{4})$ & $(240.1,1122.6,2748.7)$ & $(178.3,2670.4,3179.7)$ \tabularnewline
        \addlinespace
        $(\Tilde{t}_{1},\Tilde{t}_{2})$ & $(4631.2,4631.7)$ & $(826.2,1330.9)$  \tabularnewline
        \addlinespace
        $(\Tilde{b}_{1},\Tilde{b}_{2})$ & $(4245.0,4628.2)$ & $(1094.0,4721.6)$ \tabularnewline
        \bottomrule
        \addlinespace
    \end{tabular}
\end{table}
\begin{table}
    \caption{\label{tab:SUSY-spectrum-BRPV3}
    SUSY mass spectrum corresponding to the
    benchmark models \texttt{BM$\mu$} and \texttt{BM$\tau$}. For the states $\Tilde{\nu}_{l}^{E}$, $\Tilde{\nu}_{l}^{O}$ and $\Tilde{l}_{i}$ we have $l=\mu$ in the \texttt{BM$\mu$} model and $l=\tau$ in \texttt{BM$\tau$}. See also caption to Table~\ref{tab:SUSY-spectrum-BRPV1}.}
    \medskip{}
    \centering{}
    \begin{tabular}{ccccc}
        \toprule 
         & \texttt{BM$\mu$} & \texttt{BM$\tau$}  \tabularnewline
        \midrule
        \addlinespace
        $(\tilde\chi_{2}^{0},\tilde\chi_{3}^{0},\tilde\chi_{4}^{0},\tilde\chi_{5}^{0})$ & $(632.2,713.1,761.7,1428.1)$ & $(633.1,707.0,760.1,1653.1)$ \tabularnewline
        \addlinespace
        $(\chi_{2}^{\pm},\chi_{3}^{\pm})$ & $(633.2,764.0)$ & $(634.1,762.0)$  \tabularnewline
        \addlinespace
        $(\Tilde{\nu}_{l}^{E},H)$ & $(181.3,8996.4)$ & $(174.8,8545.1)$  \tabularnewline
        \addlinespace
        $(\Tilde{\nu}_{l}^{O},A)$ & $(181.3,8996.4)$ & $(174.8,8545.0)$  \tabularnewline
        \addlinespace
        $(\Tilde{l}_{2},\Tilde{l}_{3},\Tilde{l}_{4})$ & $(197.4,4249.4,8996.7)$ & $(191.5,4145.2,8545.4)$ \tabularnewline
        \addlinespace
        $(\Tilde{t}_{1},\Tilde{t}_{2})$ & $(2201.0,2233.7)$ & $(2416.1,2427.2)$  \tabularnewline
        \addlinespace
        $(\Tilde{b}_{1},\Tilde{b}_{2})$ & $(2210.7,4720.7)$ & $(2415.5,4594.0)$ \tabularnewline
        \bottomrule
        \addlinespace
    \end{tabular}
\end{table}
\begin{table}
    \caption{\label{tab:SUSY-spectrum-TRPV1}
    SUSY mass spectrum corresponding to the
    benchmark models \texttt{BM$\lambda^{\prime}_{333}$} and \texttt{BM$\lambda^{\prime}_{311}$}. See also caption to Table~\ref{tab:SUSY-spectrum-BRPV1}.}
    \medskip{}
    \centering{}
    \begin{tabular}{ccccc}
        \toprule 
         & \texttt{BM$\lambda^{\prime}_{333}$} & \texttt{BM$\lambda^{\prime}_{311}$}  \tabularnewline
        \midrule
        \addlinespace
        $(\tilde\chi_{2}^{0},\tilde\chi_{3}^{0},\tilde\chi_{4}^{0},\tilde\chi_{5}^{0})$ & $(150.2,205.4,303.5,762.9)$ & $(552.0,558.0,1594.1,1749.2)$ \tabularnewline
        \addlinespace
        $(\chi_{2}^{\pm},\chi_{3}^{\pm})$ & $(153.1,305.9)$ & $(554.2,1594.2)$  \tabularnewline
        \addlinespace
        $(\Tilde{\nu}_{\tau}^{E},H)$ & $(725.7,2166.1)$ & $(203.2,1661.2)$  \tabularnewline
        \addlinespace
        $(\Tilde{\nu}_{\tau}^{O},A)$ & $(725.7,2165.0)$ & $(203.1,1661.1)$  \tabularnewline
        \addlinespace
        $(\Tilde{\tau}_{2},\Tilde{\tau}_{3},\Tilde{\tau}_{4})$ & $(729.1,2166.5,2357.4)$ & $(218.2,1663.0,3693.5)$ \tabularnewline
        \addlinespace
        $(\Tilde{t}_{1},\Tilde{t}_{2})$ & $(3443.2,3487.3)$ & $(2014.4,2017.0)$  \tabularnewline
        \addlinespace
        $(\Tilde{b}_{1},\Tilde{b}_{2})$ & $(2764.4,3461.0)$ & $(2008.1,2421.6)$ \tabularnewline
        \bottomrule
        \addlinespace
    \end{tabular}
\end{table}
\begin{table}
    \caption{\label{tab:SUSY-spectrum-TRPV2}
    SUSY mass spectrum corresponding to the
    benchmark models \texttt{BM$\lambda_{233}$} and \texttt{BM$\lambda_{322}$}.
    For the states $\Tilde{\nu}_{l}^{E}$, $\Tilde{\nu}_{l}^{O}$ and $\Tilde{l}_{i}$ we have $l=\mu$ in the \texttt{BM$\lambda_{233}$} model and $l=\tau$ in \texttt{BM$\lambda_{322}$}. See also 
    caption to Table~\ref{tab:SUSY-spectrum-BRPV1}.}
    \medskip{}
    \centering{}
    \begin{tabular}{ccccc}
        \toprule 
         & \texttt{BM$\lambda_{233}$} & \texttt{BM$\lambda_{322}$}  \tabularnewline
        \midrule
        \addlinespace
        $(\chi_{2}^{0},\chi_{3}^{0},\chi_{4}^{0},\chi_{5}^{0})$ & $(589.7,951.3,959.9,1367.1)$ & $(236.3,271.8,319.4,1228.8)$ \tabularnewline
        \addlinespace
        $(\chi_{2}^{\pm},\chi_{3}^{\pm})$ & $(948.3,1367.1)$ & $(265.7,1228.8)$  \tabularnewline
        \addlinespace
        $(\Tilde{\nu}_{l}^{E},H)$ & $(207.5,2143.0)$ & $(709.3,5009.4)$  \tabularnewline
        \addlinespace
        $(\Tilde{\nu}_{l}^{O},A)$ & $(207.5,2142.5)$ & $(709.3,5008.9)$  \tabularnewline
        \addlinespace
        $(\Tilde{l}_{2},\Tilde{l}_{3},\Tilde{l}_{4})$ & $(220.9,2144.0,3133.3)$ & $(712.7,2371.3,5009.5)$ \tabularnewline
        \addlinespace
        $(\Tilde{t}_{1},\Tilde{t}_{2})$ & $(2592.6,2600.9)$ & $(2735.2,2839.2)$  \tabularnewline
        \addlinespace
        $(\Tilde{b}_{1},\Tilde{b}_{2})$ & $(2591.0,4703.4)$ & $(2381.9,2782.3)$ \tabularnewline
        \bottomrule
        \addlinespace
    \end{tabular}
\end{table}

\acknowledgments

The work of AS  was supported in part by the U.S. DOE contract \#DE-SC0012704. JC would like to thank Tal Miller for his assistance in the numerical sessions.

\newpage


\bibliographystyle{hunsrt.bst}
\bibliography{mybib2}

\end{document}